\RequirePackage{fix-cm}
\documentclass{svjour3}                     
\smartqed  
\usepackage{graphicx}
\graphicspath{{figures/}}
\usepackage{amsmath,amssymb,amsfonts}
\usepackage{subfigure}
\usepackage[numbers]{natbib}
\usepackage{algorithm}
\usepackage{algpseudocode}
\usepackage{amsbsy}
\usepackage{multirow}
\usepackage{framed} 
\usepackage{multicol} 
\usepackage{morefloats}

\begin{document}

\title{Generalized deconvolution procedure for structural modeling of turbulence} 

\titlerunning{Generalized deconvolution procedure for structural modeling of turbulence}        

\author{Omer San         \and
        Prakash Vedula 
}


\institute{O. San \at
              Mechanical \& Aerospace Engineering \\
              Oklahoma State University \\
              Tel.: +1 (405) 744-2457 \\
              \email{osan@okstate.edu}           
           \and
            P. Vedula \at
            Aerospace \& Mechanical Engineering \\
            University of Oklahoma \\
            Tel.: +1 (405) 325-4361 \\
            \email{pvedula@ou.edu}           
}

\date{Received: \today / Accepted: date}

\maketitle

\begin{abstract}
Approximate deconvolution forms a mathematical framework for the structural modeling of turbulence. The sub-filter scale flow quantities are typically recovered by using the Van Cittert iterative procedure. In this paper, however, we put forth a generalized approach for the iterative deconvolution process of sub-filter scale recovery of turbulent flows by introducing Krylov space iterative methods. Their accuracy and efficiency are demonstrated through a systematic a-priori analysis of solving the Kraichnan and Kolmogorov homogeneous isotropic turbulence problems in two- and three-dimensional domains, respectively. Our numerical assessments show that the conjugate gradient based iterative techniques lead to significantly improved performance over the Van Cittert procedure and offer great promise for approximate deconvolution turbulence models. In fact, our energy spectra analysis illustrates that a substantially longer inertial range can be recovered by using the proposed procedure equipped with the BiCGSTAB iterative scheme. This trend is also confirmed by capturing tails of the probability density function of turbulent flow quantities.

\keywords{Inverse problems \and Approximate deconvolution \and Van Cittert iterations \and Krylov space methods \and Sub-filter scale modeling \and Kraichnan turbulence \and Kolmogorov turbulence}
\end{abstract}

\section{Introduction}
\label{Intro}
Turbulent flows are encountered in a variety of engineering and geophysical systems involving a wide range of spatial and temporal scales. In a direct numerical simulation (DNS), the full spectra of turbulence should be resolved down to the Kolmogorov scale where the smallest feature of the motion is captured. The resolution requirements, however, are computationally prohibitive to fully resolve for all associated scales. On the other hand, large eddy simulation (LES) aims to reduce this computational complexity and has been proven to be a promising approach for calculations of complex turbulent flows \citep{boris1992new,lesieur1996new,piomelli1999large,meneveau2000scale}. Allowing for much coarser spatial meshes, LES is designed to resolve the most energetic large scales of the turbulent motion while modeling small scales.

In the past few decades there has been a substantial effort on developing LES closure models using physical or mathematical arguments \citep{sagaut2006large,berselli2006mathematics}. Stolz and Adams \cite{stolz1999approximate} proposed an {\em approximate deconvolution (AD)} framework to estimate the sub-filter scale quantities from the filtered flow variables. Conceptually borrowed from the image processing community \cite{biemond1990iterative}, this structural closure model utilizes {\em Van Cittert} iterations by employing repeated filtering operators to represent unfiltered small scale contributions \cite{layton2012approximate}. Therefore, the use of Van Cittert iterations in the AD process constitutes a state-of-the-art procedure for structural modeling of turbulence and sub-filter scale recovery. However, very few papers have touched on the performance of the iterative solvers for the AD procedure \citep{al2012comprehensive,maulik2016stable}, and this work aims to partially fill this gap.

The AD procedure gives an estimate for the sub-filter scale quantities to obtain an approximated LES closure model using repeated filtering operations on a computational grid. It should be pointed out that the Van Cittert iterations can be mathematically considered as Richardson iterations \cite{van2003iterative}. Although the Richardson iterative procedure constitutes a foundation for many fixed point iterative schemes, it is well known that it has a slow convergence property. On the other hand, many successful approaches based on Krylov subspace solvers have been introduced for linear solvers \cite{saad2003iterative}.

In this paper, we propose a modular iterative approach to accelerate the convergence of the AD process for structural modeling of turbulent flows. We utilize the Krylov subspace iterative methods for general AD process of the recovery of sub-filter scales. It is shown that the conjugate gradient based approaches provide significant improvement in performance over the standard Van Cittert iterations. Two test problems displaying homogeneous isotropic turbulence are chosen in our numerical assessments to illustrate the success of the proposed iterative approaches in our generalized AD framework.

\section{Closure modeling}
\label{s:sfs}

To illustrate the LES framework we first consider the incompressible Navier-Stokes equations in dimensionless conservative form \cite{kim1985application}
\begin{align} \label{eq:ge}
 \frac{\partial u_i}{\partial t} + \frac{\partial (u_i u_j)}{\partial x_j} = -\frac{\partial p}{\partial x_i} + \frac{1}{Re} \frac{\partial^2 u_i}{\partial x_j \partial x_j},
\end{align}
where we use the incompressibility constraints as
\begin{align} \label{eq:comp}
 \frac{\partial u_j}{\partial x_j} =0,
\end{align}
and $u_i$ are the velocity components, $p$ is the pressure, and $Re$ is the Reynolds number. For LES computations, the governing equations are filtered in space and solved numerically on a grid, hence the filtered equations of motion can be obtained by performing a low-pass filtering in the following form
\begin{align} \label{eq:fge}
 \frac{\partial \bar{u}_i}{\partial t} + \frac{\partial (\overline{u_i u_j})}{\partial x_j} = -\frac{\partial \bar{p}}{\partial x_i} + \frac{1}{Re} \frac{\partial^2 \bar{u}_i}{\partial x_j \partial x_j},
\end{align}
where $\bar{u}_i$ denotes the filtered velocity components on an LES grid. The effective low-pass filtered LES equation can be rewritten as
\begin{align} \label{eq:fge2}
 \frac{\partial \bar{u}_i}{\partial t} + \frac{\partial (\bar{u}_i \bar{u}_j)}{\partial x_j} = -\frac{\partial \bar{p}}{\partial x_i} + \frac{1}{Re} \frac{\partial^2 \bar{u}_i}{\partial x_j \partial x_j} + \frac{\partial \tau_{ij}  }{\partial x_j} ,
\end{align}
where the turbulent stresses (i.e., so-called Reynolds stresses) become
\begin{align} \label{eq:fge3}
\tau_{ij} = \bar{u}_i \bar{u}_j - \overline{u_i u_j},
\end{align}
in which the last term constitutes the closure problem for the LES computations.

More precisely, the solution of Eq.~(\ref{eq:fge2}) returns the filtered quantities of $\bar{u}_i$, however, an estimate to the unclosed term $\overline{u_i u_j}$ in Eq.~(\ref{eq:fge3}) should be provided to account for nonlinear interactions at sub-filter scales. If these inter-eddy nonlinear interactions are not properly parameterized, then an increase in resolution will not necessarily improve the accuracy of these large scales \cite{frederiksen2016theoretical}. Therefore, a central challenge in turbulence simulations is to achieve an accurate closure of these coarse-grained LES equations by relating unclosed terms to resolved quantities via appropriate turbulence models. Many such models have been proposed with varying levels of sophistication \cite{sagaut2006large}. In this study, we focus on a structural approach to turbulence closure problem where an iterative deconvolution procedure is used to establish a relationship between unclosed terms and resolved variables.

\section{Generalized approximate deconvolution framework}
\label{s:gad}

The idea of spatial filtering is central in LES. The most popular LES modeling approach is known as the \emph{functional approach} to artificial eddy viscosity parametrization where the assumption of small scale isotropy is utilized to represent the universal characteristic of the dissipation of the scales in the dissipation range \cite{frisch1995turbulence}.
On the other hand, \emph{structural approaches} address the closure problem by extrapolating from the resolved scales to the unresolved scales. A popular closure approach is scale similarity, introduced by Bardina, Ferziger and Reynolds \citep{bardina1980improved}
\begin{align} \label{eq:tau1}
\tau_{ij} = \bar{\bar{u}}_i \bar{\bar{u}}_j - \overline{\bar{u}_i \bar{u}_j},
\end{align}
which has been proven accurate in a-priori setting \citep{sarghini1999scale}. As highlighted in \citep{layton2003simple}, the scale similarity models are reversible; they can provide so-called backscatter; and they best align with the principal axis of the true stress tensor; however, they show stability problems when used in a numerical simulation. A simple yet stable model introduced by Layton and Lewandowski \citep{layton2003simple} reads as
\begin{align} \label{eq:tau2}
\tau_{ij} = \bar{u}_i \bar{u}_j - \overline{\bar{u}_i \bar{u}_j},
\end{align}
which has a superior stability property. A generalization of this model yields a family of structural models, known as approximate deconvolution (AD). The AD framework offer an iterative structural modeling approach which employs repeated filtering of the filtered variables to obtain an approximation of the unfiltered variables. The basic problem in the AD framework: solve a general equation $\bar{u}=Gu$ where $G$ is a filtering operator, $\bar{u}$ is a known quantity and $u$ is a solution quantity to be determined. In other words, it reads to solve the equation below iteratively \citep{layton2007similarity}
\begin{align} \label{eq:tau3}
Gu = \bar{u}, \textnormal{   solve for $u$.}
\end{align}
The filtering equation under consideration can also be expressed as $u=G^{-1}\bar{u}$, where $G^{-1}$ denotes the filtering operator inverse. Since the low-pass filter $G=I-(I-G)$, this equation can also be written in an alternate form as
\begin{eqnarray}
u &=& [{I-(I-G)}]^{-1}\bar{u} \\
u &=& [I + (I-G) + (I-G)^2 + (I-G)^3 + \ldots]\bar{u}
\end{eqnarray}
where $I$ denotes the identity operator. Alternatively, an inverse to $G$ can be written formally as the Neumann series
\begin{equation}
    G^{-1} = \sum_{n=0}^{N-1}(I-G)^{n},
\end{equation}
where $G^{-1}$ will asymptotically approach the true inverse of the filtering operator $G$ as $N \rightarrow \infty$, provided the series convergences. An iterative solution approach based on deconvolution can be written as
\begin{eqnarray}\label{eq:adm1}
    u_{0} &=& \bar{u} \nonumber\\
    u_{n+1} &=& \bar{u} + (I - G)u_{n}, \quad n = 0,1,2,...,N-1.
\end{eqnarray}
This approach was studied by Van Cittert and its use in LES was pioneered by Stolz and Adams \citep{stolz1999approximate}. The convergence of this solution approach depends on the spectral radius of the operator $I-G$. Since most of the filtering operators have their transfer functions between $0 \leq \hat{G}(k) \leq 1$, the solution will converge. In other words, this procedure is
numerically stable for the following condition
\begin{align} \label{eq:radius}
|1-\hat{G}(k)| \leq 1,
\end{align}
where $\hat{G}(k)$ is the transfer function of the filtering operator $G$. In a mathematically equivalent form, this process computes an approximate solution $u$ to the above deconvolution equation by $N$ steps of a fixed point iteration problem \citep{bertero1998introduction} and can be rewritten as
\begin{eqnarray}\label{eq:adm2}
    u_{0} &=& \bar{u} \nonumber\\
    u_{n+1} &=& u_{n} + (\bar{u} - G u_{n}), \quad n = 0,1,2,...,N-1,
\end{eqnarray}
where it can be also referred to as the Richardson iterative procedure \citep{van2003iterative}. If we define the residual
\begin{align} \label{eq:res}
r = \bar{u} - Gu,
\end{align}
then we may write Eq.~(\ref{eq:adm2}) as
\begin{align} \label{eq:ric}
u_{n+1} = u_{n} + r_{n}
\end{align}
where $r_{n}$ constitutes the update between two successive iterations. We note that Eq.~(\ref{eq:tau2}) can be recovered by $N=1$, and for example, a second-order approximation can be written as
\begin{align} \label{eq:tau3}
\tau_{ij} = \bar{u}_i \bar{u}_j - \overline{(2\bar{u}_i-\bar{\bar{u}}_i) (2\bar{u}_j-\bar{\bar{u}}_j)},
\end{align}
for $N=2$ (e.g., see also \citep{germano2015similarity}). The general procedure for the Van Cittert approximate deconvolution process is given in Algorithm~\ref{alg:vc}. We refer readers to the monograph by Layton and Rebholz \cite{layton2012approximate} on further discussion of the AD models.

\begin{algorithm}[!h]
\caption{Van Cittert approximate deconvolution algorithm}\label{alg:vc}
\begin{algorithmic}[1]
\State Given $\bar{\boldsymbol u}$ \Comment{Given filtered data in space, i.e., $\bar{\boldsymbol u} = \bar{u}(\mathbf{x})$ }
\State Given filtering operator $G$  \Comment{$\bar{\boldsymbol u} = G \boldsymbol u$}
\State $\boldsymbol u_0 = \bar{\boldsymbol u}$ \Comment{Initialize solution}
\State $\boldsymbol r_0= \bar{\boldsymbol u} - G \boldsymbol u_{0}$ \Comment{Initialize residual}
\State $n=0$
\While{$n < N$}
\State $\boldsymbol u_{n+1} = \boldsymbol u_{n} + \boldsymbol r_n$ \Comment{$\boldsymbol u_{n+1} = \boldsymbol u_{n} + (\bar{\boldsymbol u} - G \boldsymbol u_{n})$}
\State $\boldsymbol d_n = G \boldsymbol r_n$ \Comment{Apply filtering operator}
\State $\boldsymbol r_{n+1} = \boldsymbol r_{n} - \boldsymbol d_n$ \Comment{$\boldsymbol r_{n+1} = \bar{\boldsymbol u} - G \boldsymbol u_{n+1}$}
\State $n \leftarrow n+1$
\EndWhile
\end{algorithmic}
\end{algorithm}

The van Cittert method has likely become the most common technique for deconvolution due to its following advantages; it is simple to practically use; easy to analyze mathematically, convergent and has performed well in many computations \citep{layton2012approximate}, and typically $N$ is kept small $N\leq5$.
The primary motivation for the current study is to accelerate the convergence rate of the deconvolution procedure. We highlight that the Van Cittert iterative process (i.e., also known as the Richardson or Jacobi iterations) results in a slow convergence rate and deteriorates in performance as the degrees of freedom of the linear system increases \citep{van2003iterative}. According to the authors' best knowledge, however, there has been no investigation for the acceleration of the AD procedure for sub-filter scale recovery. To improve its convergence rate further, instead of a fixed step size in each iteration, we introduce a dynamic distance parameter $\alpha_n$ in the direction of steepest increase as follows
\begin{align} \label{eq:sd}
u_{n+1} = u_{n} + \alpha_n r_{n}.
\end{align}
The value of $\alpha_n$ can be computed satisfying the orthogonality of the subsequent search directions, i.e.,
\begin{align} \label{eq:inner}
\int r_{n}(\mathbf{x}) r_{n+1}(\mathbf{x})d\mathbf{x} = 0,
\end{align}
where $\mathbf{x}$ is the space where we define solution data $u$ and its associated residual $r$. This inner product can be written as
\begin{align} \label{eq:inner2}
(r_{n},r_{n+1}) = 0.
\end{align}
Using the definitions given by Eq.~(\ref{eq:res}) and Eq.~(\ref{eq:sd}) we may compute the residual as follows
\begin{eqnarray}\label{eq:rnew1}
    r_{n+1} &=& \bar{u} -G u_{n+1} \nonumber\\
            &=& \bar{u} -G (u_{n} + \alpha_n r_{n}) \nonumber\\
            &=& r_{n} -\alpha_n Gr_{n}.
\end{eqnarray}
Now we can easily calculate the distance parameter $\alpha_n$ to satisfy the above orthogonality condition given by Eq.~(\ref{eq:inner2})
\begin{align} \label{eq:inner3}
(r_{n},r_{n} -\alpha_n Gr_{n}) = 0,
\end{align}
which leads to
\begin{align} \label{eq:sd-a}
\alpha_n=\frac{(r_{n},r_{n})}{(r_{n},Gr_{n})}.
\end{align}
This procedure is called the \emph{method of steepest descent}. A pseudocode algorithm of this method appears in Algorithm~\ref{alg:sd} for the AD process. Successive iterations in the steepest descent method tend to oscillate back and forth towards near the solution. Thus, the remedy for the associated slow convergence is to choose other search directions. With the celebrated conjugate gradient method, we choose
\begin{align} \label{eq:cg1}
u_{n+1} = u_{n} + \alpha_n p_{n},
\end{align}
with
\begin{align} \label{eq:cg2}
p_{n} = r_{n} + \lambda (u_{n}-u_{n-1}),
\end{align}
where the new search direction is a linear combination of the steepest descent direction and the previous step correction. Using Eq.~(\ref{eq:cg1}) we can write Eq.~(\ref{eq:cg2}) as
\begin{align} \label{eq:cg3}
p_{n} = r_{n} + \lambda \alpha_{n-1} p_{n-1},
\end{align}
or rewrite with definition of a new parameter $\beta_{n-1}=\lambda \alpha_{n-1}$
\begin{align} \label{eq:cg4}
p_{n+1} = r_{n+1} + \beta_{n} p_{n},
\end{align}
where the parameters $\alpha_n$ and $\beta_{n}$ are to be determined so that convergence is as fast as possible. As with the steepest descent, we choose these parameters to satisfy conjugate orthogonality conditions between subsequent iterates. If the search directions are conjugate
\begin{align} \label{eq:cg5}
(r_{n},r_{n+1}) = 0,
\end{align}
\begin{align} \label{eq:cg6}
(p_{n},r_{n+1}) = 0.
\end{align}
Furthermore, using an inner product of Eq.~(\ref{eq:cg4}) with $r_{n+1}$ yields
\begin{align} \label{eq:cg7}
(p_{n+1},r_{n+1}) = (r_{n+1},r_{n+1}) + \beta_{n} (p_{n},r_{n+1}),
\end{align}
which can be further simplified to
\begin{align} \label{eq:cg8}
(p_{n+1},r_{n+1}) = (r_{n+1},r_{n+1}).
\end{align}
To find $\alpha_n$, using Eq.~(\ref{eq:cg1}) we first write residual
\begin{eqnarray}\label{eq:rnew2}
    r_{n+1} &=& \bar{u} -G u_{n+1} \nonumber\\
            &=& \bar{u} -G (u_{n} + \alpha_n p_{n}) \nonumber\\
            &=& r_{n} -\alpha_n Gp_{n}.
\end{eqnarray}
and then project Eq.~(\ref{eq:rnew2}) to the $p_n$ direction to obtain
\begin{align} \label{eq:cg9}
(p_{n},r_{n+1}) = (p_{n},r_{n}) - \alpha_n(p_{n},Gp_{n}),
\end{align}
which leads to
\begin{align} \label{eq:cg-a}
\alpha_n=\frac{(p_{n},r_{n})}{(p_{n},Gp_{n})}.
\end{align}
If we select $p_0=r_0$ and Eq.~(\ref{eq:cg8}) reduces to,
\begin{align} \label{eq:cg88}
(p_{n},r_{n}) = (r_{n},r_{n}),
\end{align}
then we may write the above expression as follows
\begin{align} \label{eq:cg-alpha}
\alpha_n=\frac{(r_{n},r_{n})}{(p_{n},Gp_{n})}.
\end{align}
In a similar manner, if we project Eq.~(\ref{eq:cg4}) to the $r_n$ direction
\begin{align} \label{eq:cg10}
(r_{n},p_{n+1}) = (r_{n},r_{n+1}) + \beta_n(r_{n},p_{n}),
\end{align}
and using Eq.~(\ref{eq:cg5}), the parameter $\beta_n$ can be written as
\begin{align} \label{eq:cg-b}
\beta_n=\frac{(r_{n},p_{n+1})}{(r_{n},p_{n})}.
\end{align}
If the search directions are orthogonal
\begin{align} \label{eq:cg11}
(p_{n+1},Gp_{n}) = 0,
\end{align}
it corresponds to the following definition
\begin{align} \label{eq:cg12}
(r_{n},p_{n+1}) = (r_{n+1},p_{n+1}) = (r_{n+1},r_{n+1}).
\end{align}
Using the above identity and the identity given by Eq.~(\ref{eq:cg88}), $\beta_n$ can be rewritten as
\begin{align} \label{eq:cg-b}
\beta_n=\frac{(r_{n+1},r_{n+1})}{(r_{n},r_{n})}.
\end{align}
This completes the derivation of the conjugate gradient (CG) method illustrated in Algorithm~\ref{alg:cg}. Furthermore, Algorithm~\ref{alg:bi} demonstrates the biconjugate gradient stabilized method, often abbreviated as BiCGSTAB \citep{van1992bi}, for the approximate deconvolution procedure. This algorithm can also be considered the generalized minimal residual method (usually abbreviated as GMRES \citep{saad2003iterative}) with deflated restarting (i.e., GMRES(1) since we restart from initial residual).

\begin{algorithm}[!h]
\caption{Steepest descent approximate deconvolution algorithm}\label{alg:sd}
\begin{algorithmic}[1]
\State Given $\bar{\boldsymbol u}$ \Comment{Given filtered data in space, i.e., $\bar{\boldsymbol u} = \bar{u}(\mathbf{x})$}
\State Given filtering operator $G$  \Comment{$\bar{\boldsymbol u} = G \boldsymbol u$}
\State $\boldsymbol u_0 = \bar{\boldsymbol u}$ \Comment{Initialize solution}
\State $\boldsymbol r_0= \bar{\boldsymbol u} - G \boldsymbol u_{0}$ \Comment{Initialize residual}
\State $n=0$
\While{$n < N$}
\State $\boldsymbol d_n = G \boldsymbol r_n$ \Comment{Apply filtering operator}
\State $\alpha_n = (\boldsymbol r_n, \boldsymbol r_n)/(\boldsymbol r_n, \boldsymbol d_n)$ \Comment{Inner products $(\boldsymbol a, \boldsymbol b)=\int a(\mathbf{x}) b(\mathbf{x}) d\mathbf{x}$}
\State $\boldsymbol u_{n+1} = \boldsymbol u_{n} + \alpha_n \boldsymbol r_n$ \Comment{Update solution}
\State $\boldsymbol r_{n+1} = \boldsymbol r_{n} - \alpha_n \boldsymbol d_n$ \Comment{$\boldsymbol r_{n+1} = \bar{\boldsymbol u} - G \boldsymbol u_{n+1}$}
\State $n \leftarrow n+1$
\EndWhile
\end{algorithmic}
\end{algorithm}

\begin{algorithm}
\caption{Conjugate gradient approximate deconvolution algorithm}\label{alg:cg}
\begin{algorithmic}[1]
\State Given $\bar{\boldsymbol u}$ \Comment{Given filtered data in space, i.e., $\bar{\boldsymbol u} = \bar{u}(\mathbf{x})$}
\State Given filtering operator $G$  \Comment{$\bar{\boldsymbol u} = G \boldsymbol u$}
\State $\boldsymbol u_0 = \bar{\boldsymbol u}$ \Comment{Initialize solution}
\State $\boldsymbol r_0= \bar{\boldsymbol u} - G \boldsymbol u_{0}$ \Comment{Initialize residual}
\State $\boldsymbol p_0= \boldsymbol r_0$ \Comment{Initialize conjugate}
\State $\rho_0 = (\boldsymbol r_0, \boldsymbol r_0)$
\State $n=0$
\While{$n < N$}
\State $\boldsymbol d_n = G \boldsymbol p_n$ \Comment{Apply filtering operator}
\State $\alpha_n = \rho_n/(\boldsymbol p_n, \boldsymbol d_n)$ \Comment{Inner product $(\boldsymbol a, \boldsymbol b)=\int a(\mathbf{x}) b(\mathbf{x}) d\mathbf{x}$}
\State $\boldsymbol u_{n+1} = \boldsymbol u_{n} + \alpha_n \boldsymbol p_n$ \Comment{Update solution}
\State $\boldsymbol r_{n+1} = \boldsymbol r_{n} - \alpha_n \boldsymbol d_n$ \Comment{$\boldsymbol r_{n+1} = \bar{\boldsymbol u} - G \boldsymbol u_{n+1}$}
\State $\rho_{n+1} = (\boldsymbol r_{n+1}, \boldsymbol r_{n+1})$ \Comment{Inner product $(\boldsymbol a, \boldsymbol b)=\int a(\mathbf{x}) b(\mathbf{x}) d\mathbf{x}$}
\State $\beta_n = \rho_{n+1}/\rho_{n}$
\State $\boldsymbol p_{n+1} = \boldsymbol r_{n+1} + \beta_n \boldsymbol p_n$
\State $n \leftarrow n+1$
\EndWhile
\end{algorithmic}
\end{algorithm}

\begin{algorithm}
\caption{BiCGSTAB approximate deconvolution algorithm}\label{alg:bi}
\begin{algorithmic}[1]
\State Given $\bar{\boldsymbol u}$ \Comment{Given filtered data in space, i.e., $\bar{\boldsymbol u} = \bar{u}(\mathbf{x})$}
\State Given filtering operator $G$  \Comment{$\bar{\boldsymbol u} = G \boldsymbol u$}
\State $\boldsymbol u_0 = \bar{\boldsymbol u}$ \Comment{Initialize recovered data}
\State $\boldsymbol r_0= \bar{\boldsymbol u} - G \boldsymbol u_{0}$ \Comment{Initialize residual}
\State $\boldsymbol f_0= \boldsymbol r_0$ \Comment{Frozen initial residual, i.e., GMRES(1)}
\State $\boldsymbol p_0= 0.0 $, $\boldsymbol q_0= 0.0$  \Comment{Initialize bi-conjugates}
\State $\alpha_0 = 1.0$, $\omega_0 = 1.0$, $\rho_0 = 1.0$   \Comment{Initialize constants}
\State $n=0$
\While{$n < N$}
\State $\rho_{n+1} = (\boldsymbol f_0, \boldsymbol r_n) $ \Comment{Inner product $(\boldsymbol a, \boldsymbol b)=\int a(\mathbf{x}) b(\mathbf{x}) d\mathbf{x}$}
\State $\beta_n = (\alpha_n /\omega_n )(\rho_{n+1}/ \rho_n )$
\State $\boldsymbol p_{n+1} = \boldsymbol r_{n} + \beta_n (\boldsymbol p_n - \omega_n \boldsymbol q_n)$
\State $\boldsymbol q_{n+1} = G\boldsymbol p_{n+1}$ \Comment{Apply filtering operator}
\State $\alpha_{n+1} = \rho_{n+1}/(\boldsymbol f_0, \boldsymbol q_{n+1}) $ \Comment{Inner product $(\boldsymbol a, \boldsymbol b)=\int a(\mathbf{x}) b(\mathbf{x}) d\mathbf{x}$}
\State $\boldsymbol s_{n} = \boldsymbol r_{n} - \alpha_{n+1} \boldsymbol q_{n+1}$
\State $\boldsymbol t_{n} = G \boldsymbol s_{n}$ \Comment{Apply filtering operator}
\State $\omega_{n+1} = (\boldsymbol s_{n}, \boldsymbol t_{n})/(\boldsymbol t_{n}, \boldsymbol t_{n})$ \Comment{Inner products $(\boldsymbol a, \boldsymbol b)=\int a(\mathbf{x}) b(\mathbf{x}) d\mathbf{x}$}
\State $\boldsymbol u_{n+1} = \boldsymbol u_{n} + \alpha_{n+1} \boldsymbol p_{n+1} + \omega_{n+1} \boldsymbol s_{n}$ \Comment{Update solution}
\State $\boldsymbol r_{n+1} = \boldsymbol s_{n} - \omega_{n+1} \boldsymbol t_{n}$
\State $n \leftarrow n+1$
\EndWhile
\end{algorithmic}
\end{algorithm}

\section{Results}
\label{s:results}
In this section, we present our a-priori results for both two- and three-dimensional test cases. Since filtering operator $G$ is the free modelling parameter in LES computations, to complete our discussion we choose a differential filter \citep{germano1986differential,germano1986differential_2}, which has been extensively used in LES computations (i.e., see \citep{layton2012approximate}). With the definition of operator $G=(I-\delta^2 \nabla^2)^{-1}$, the elliptic differential filter can be written as
\begin{align} \label{eq:df}
(I-\delta^2 \nabla^2)^{-1} u = \bar{u}
\end{align}
where the filtering parameter $\delta=1/k_c$ (i.e., the filter radius) controls the filter's attenuation \citep{san2016analysis}. Here $k_c$ is referred to the filter cut-off frequency, which controls the cut-off wavenumber of the filter. We parameterize $k_c=\gamma k_m$ in our computations, where $\gamma$ is a scaling parameter to control the filtering strength (i.e., $0<\gamma\leq1$). To eliminate discretization error when computing the Laplacian, we prefer to utilize a spectral method using a standard fast Fourier transform algorithm \citep{san2012high}. We note that the grid cut-off wavenumber $k_m$ corresponds to the highest wavenumber that can be resolved on a given computational domain. The transfer function of the Germano's elliptic filter is given by
\begin{equation} \label{eq:df2}
\displaystyle \hat{G}(k) = \frac{1}{1+(k/k_c)^2}.
\end{equation}
We noted that this filter is also stable by satisfying Eq.~(\ref{eq:radius}) for all wavenumbers.

\subsection{Two-dimensional Kraichnan turbulence}
\label{s:2D}

Two-dimensional Kraichnan turbulence is an homogenous isotropic decaying incompressible flow problem in which the kinetic energy decays. In this study, we solve this problem to examine the accuracy and efficiency of the presented approximate deconvolution algorithms. The two-dimensional square domain is given by a side length of $2\pi$ with periodic boundary conditions in both directions. Turbulent eddies evolve from an initial energy spectrum that decays through time. Details of our numerical methods can be found in \citep{san2012high}. A Reynolds number of $Re=32,000$ is utilized to ensure a prominent inertial range with its associated $k^{-3}$ scaling for two-dimensional turbulence predicted by the Kraichnan-Bachelor-Leith theory \citep{kraichnan1967inertial,batchelor1969computation,leith1971atmospheric}. Once the turbulence is developed, we store an instantaneous field snapshot and perform our a-priori analysis on that field data. We perform our analysis using both high and coarse resolutions using $2048^2$ and $256^2$ grid points, respectively. Coarse data are obtained from high resolution data snapshot by using a simple coarse-injection approach. From these data sets we first obtain filtered field data sets by applying the differential filter given by Eq.~(\ref{eq:df}). Starting from these filtered data sets the goal is then to recover back the true field data sets by deconvolution. Our criteria to assess the performance of proposed iterative deconvolution algorithms include computational efficiency, residual norm, energy spectrum recovery, the recovery of given turbulent field and its associated probability density function. The energy spectra are compared to the ideal $k^{-3}$ scaling as well.

\begin{figure}[!ht]
\centering
\mbox{
\subfigure[True]{\includegraphics[width=0.5\textwidth]{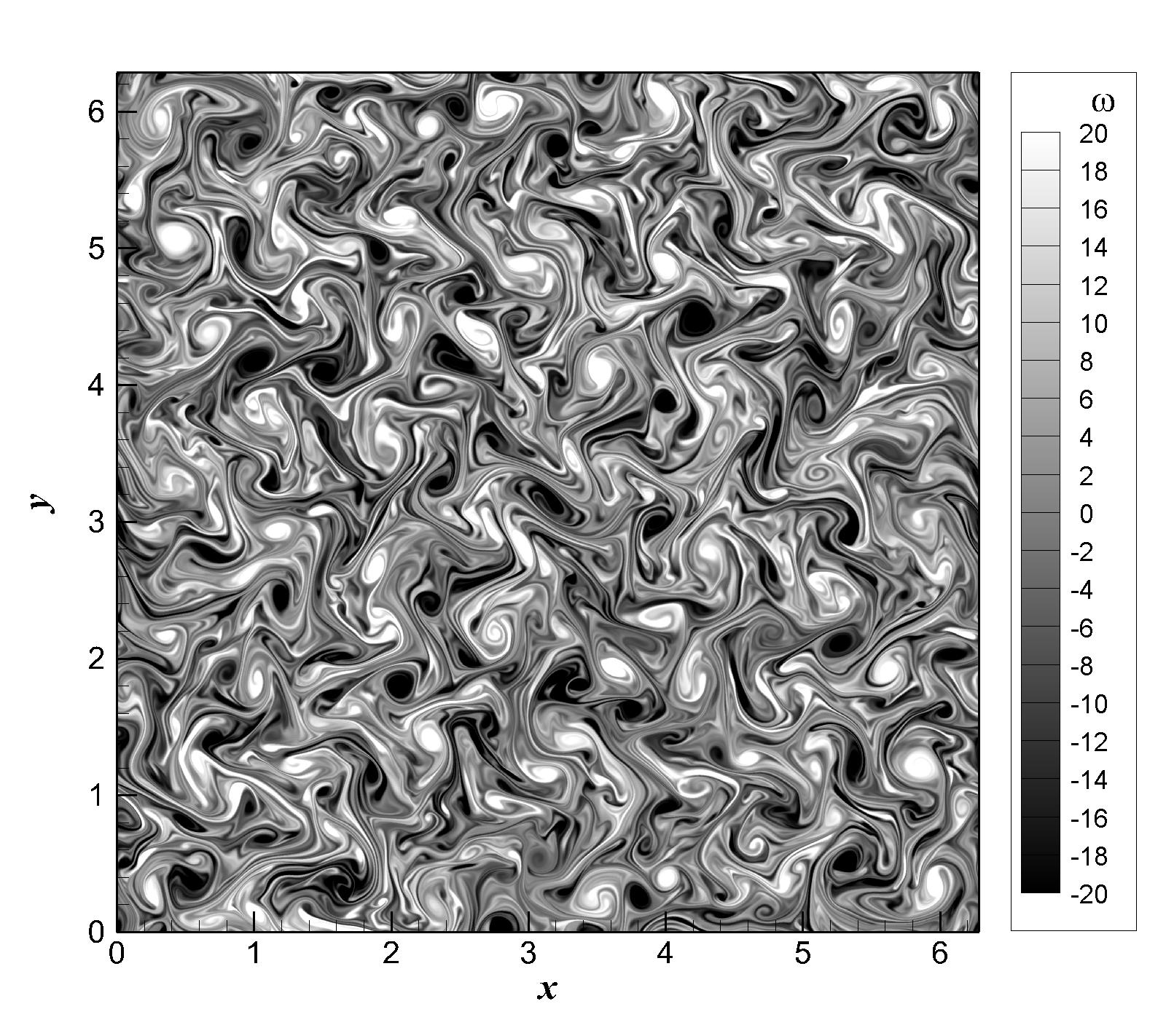}}
\subfigure[Filtered]{\includegraphics[width=0.5\textwidth]{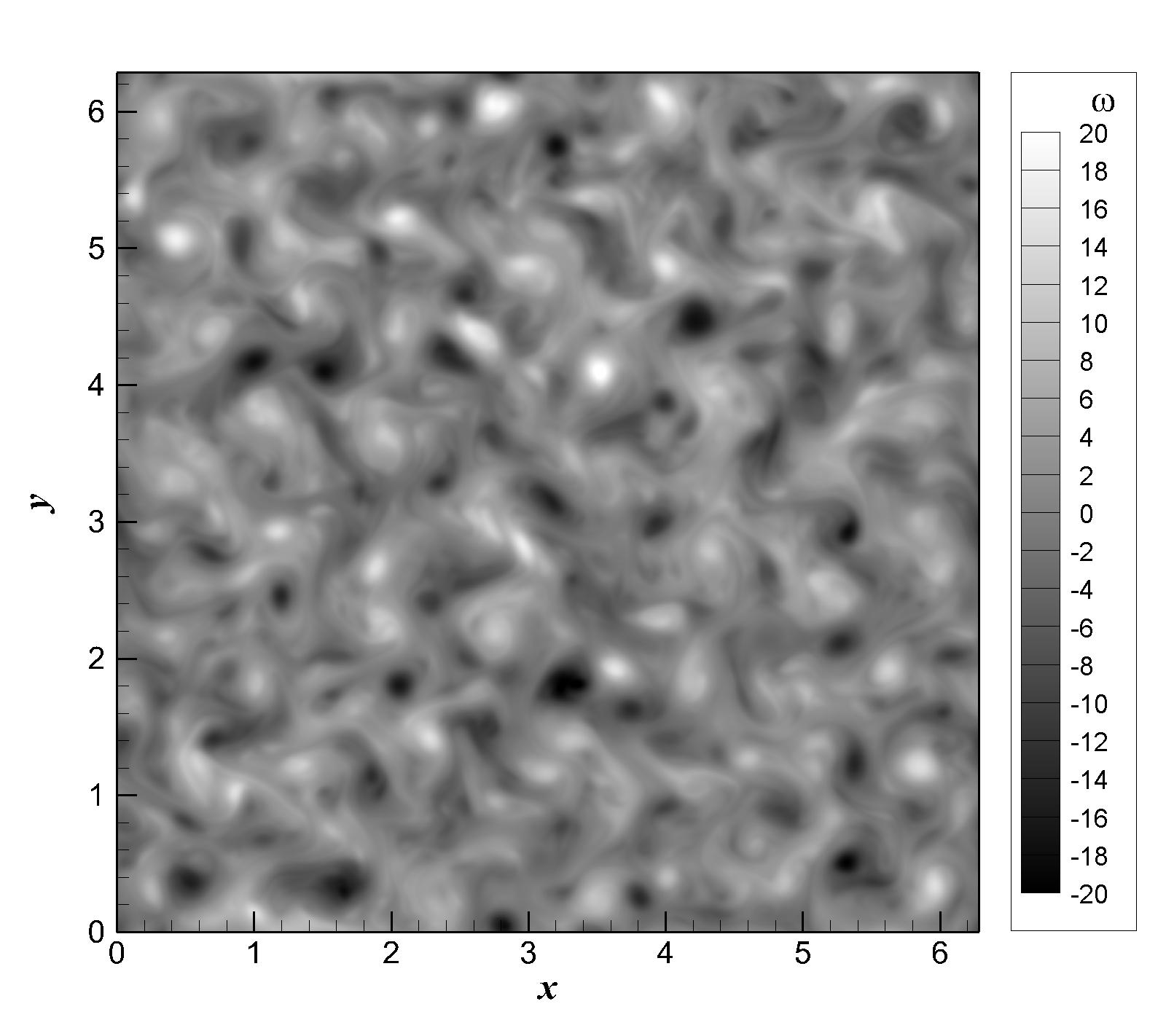}}
}\\
\mbox{
\subfigure[Van Cittert]{\includegraphics[width=0.5\textwidth]{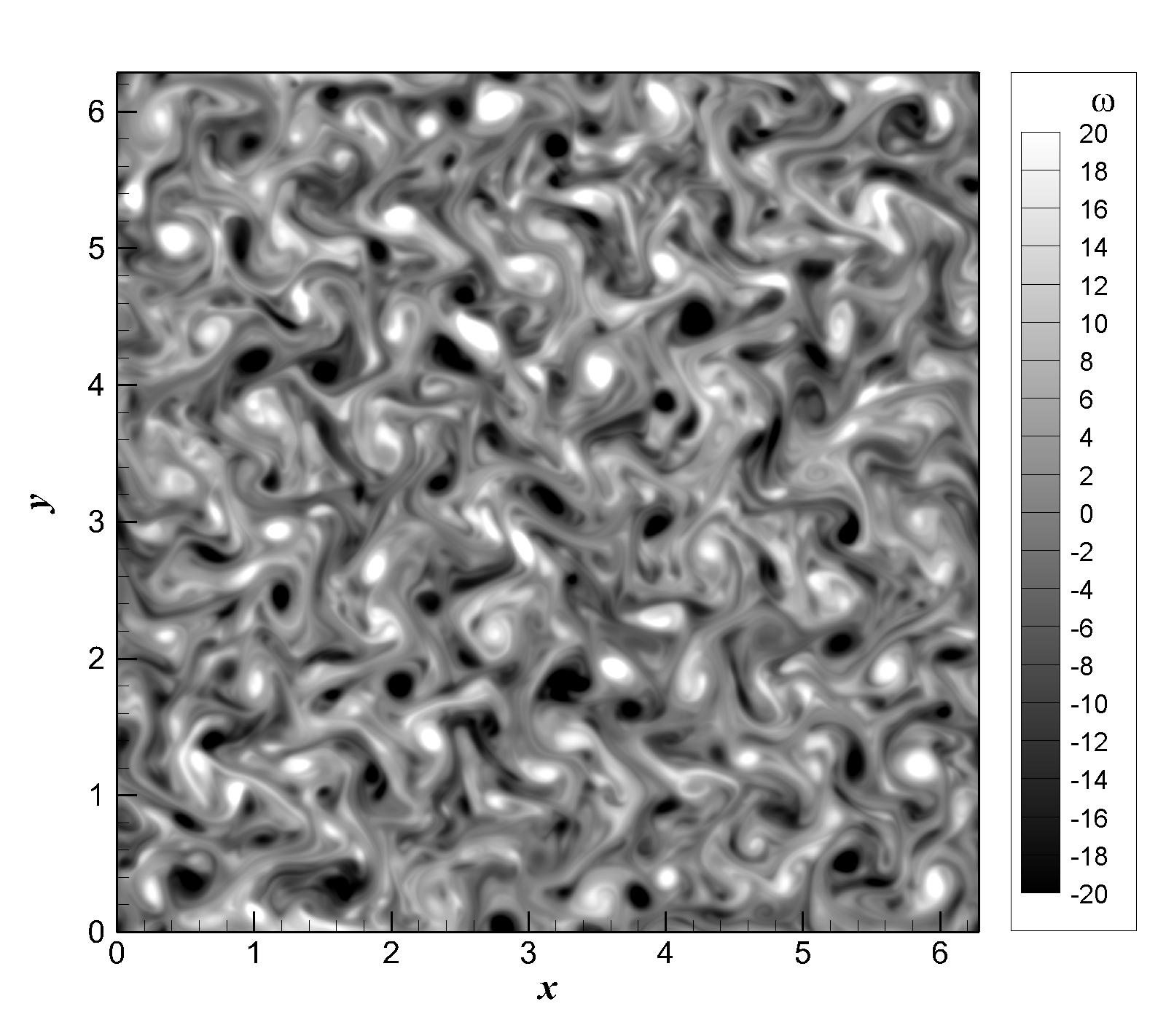}}
\subfigure[Steepest Descent]{\includegraphics[width=0.5\textwidth]{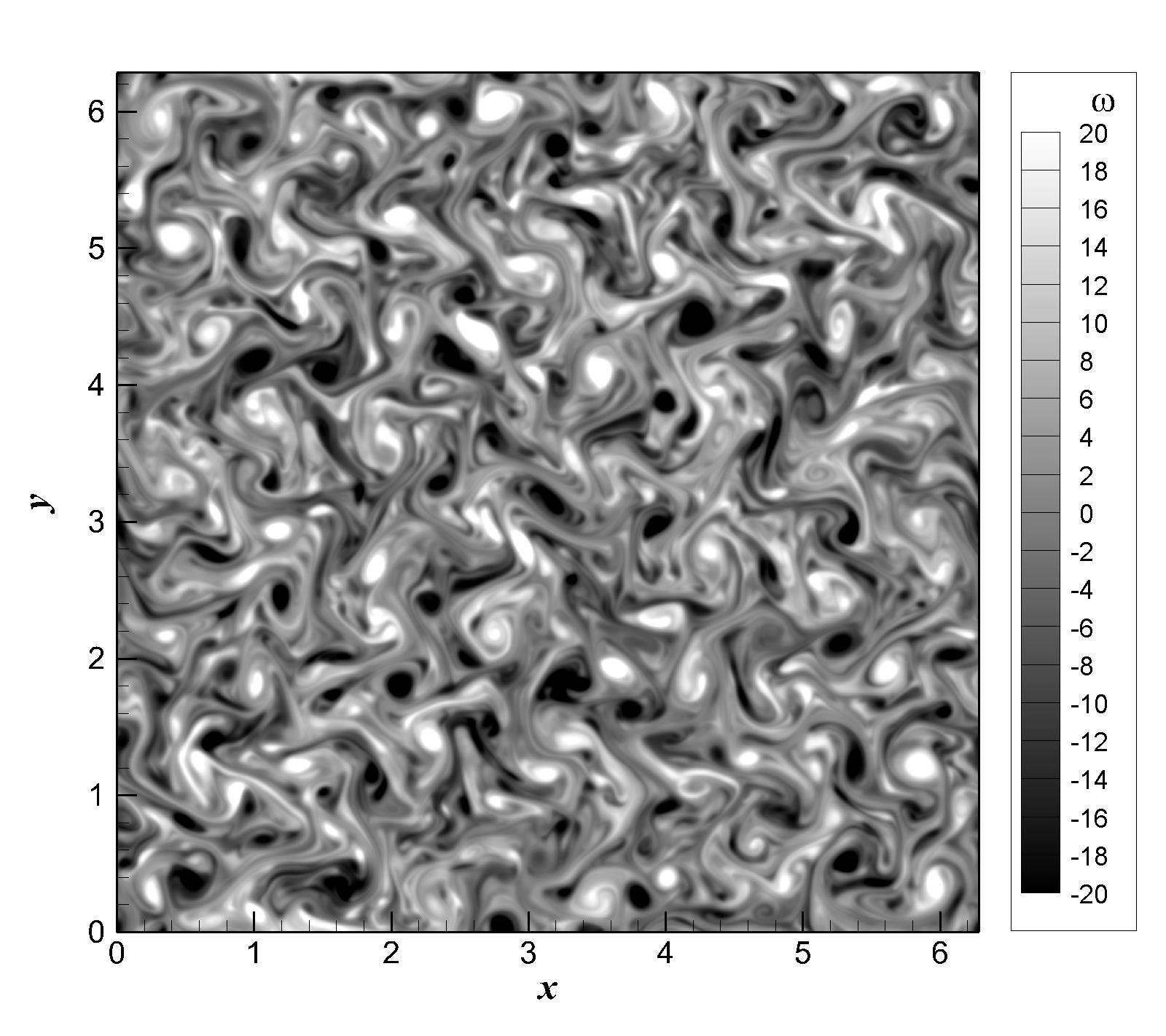}}
}\\
\mbox{
\subfigure[Conjugate Gradient]{\includegraphics[width=0.5\textwidth]{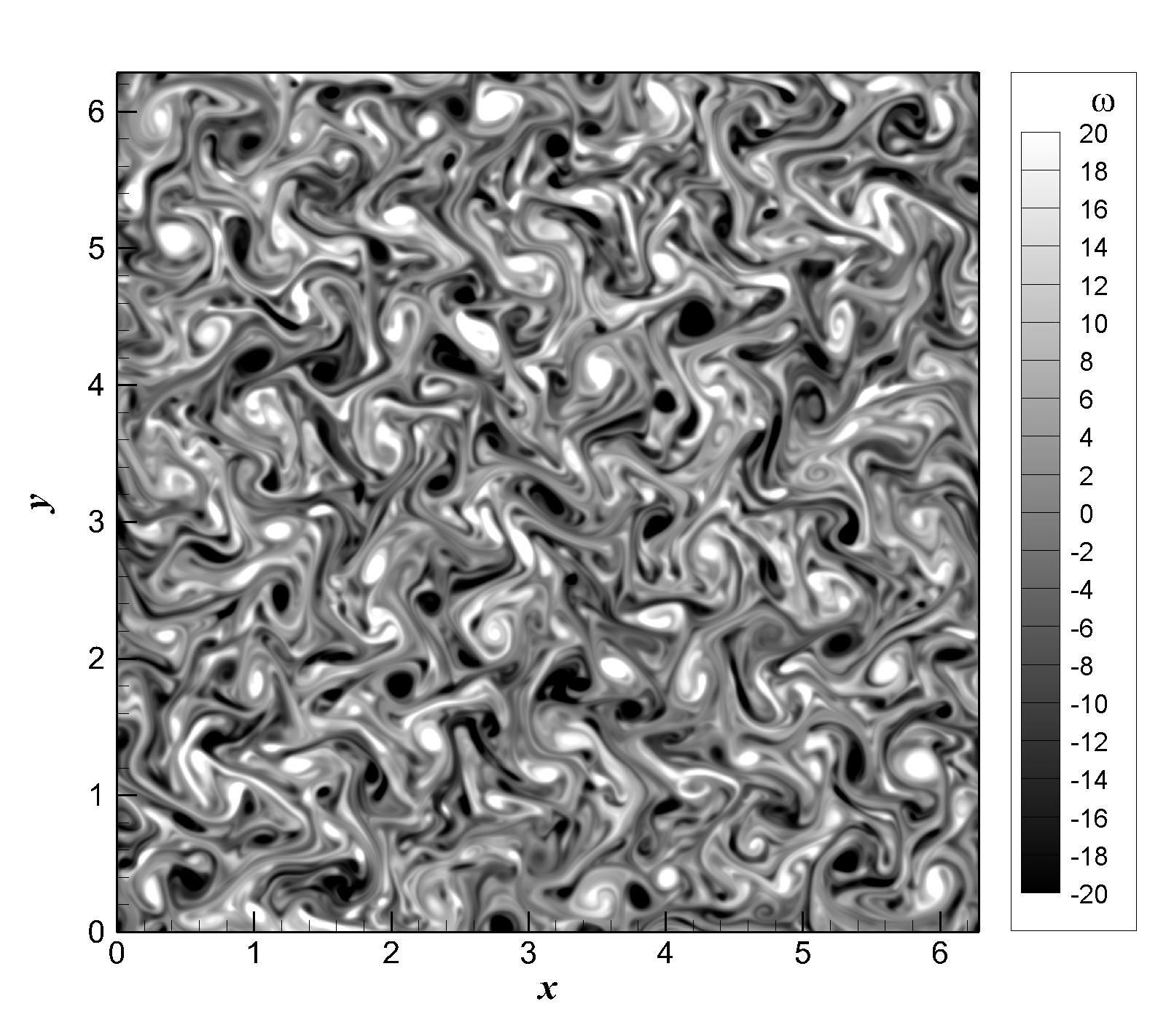}}
\subfigure[BiCGSTAB]{\includegraphics[width=0.5\textwidth]{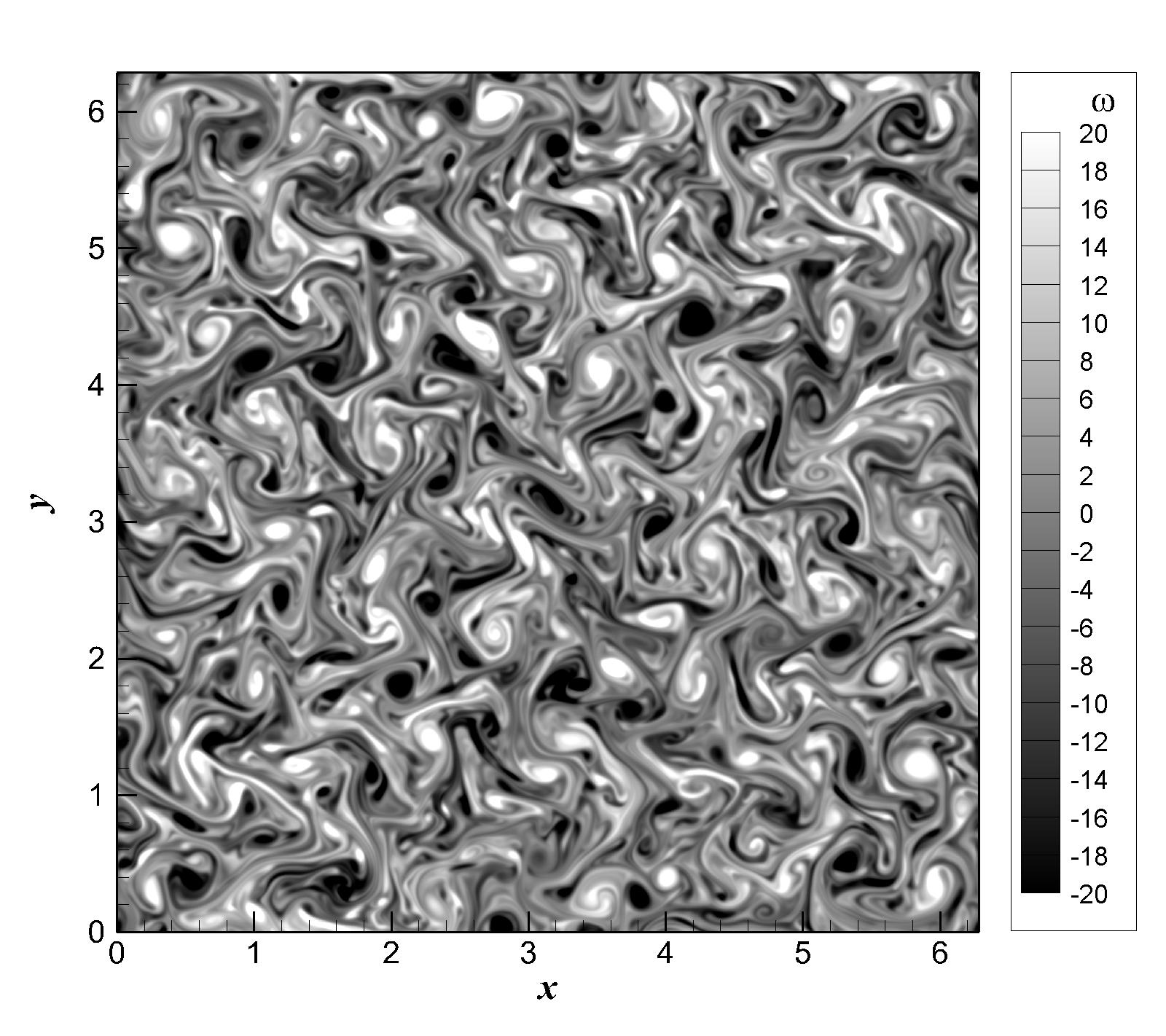}}
}
\caption{Vorticity contour fields for the two-dimensional turbulence data at $Re=32,000$ to illustrate a-priori recovery process of the true data from the filtered data using $N=5$ iterations.}
\label{f:2dfield5}
\end{figure}

\begin{figure}[!ht]
\centering
\mbox{
\subfigure[True]{\includegraphics[width=0.5\textwidth]{f_full.jpeg}}
\subfigure[Filtered]{\includegraphics[width=0.5\textwidth]{f_full_df.jpeg}}
}\\
\mbox{
\subfigure[$N=2$]{\includegraphics[width=0.5\textwidth]{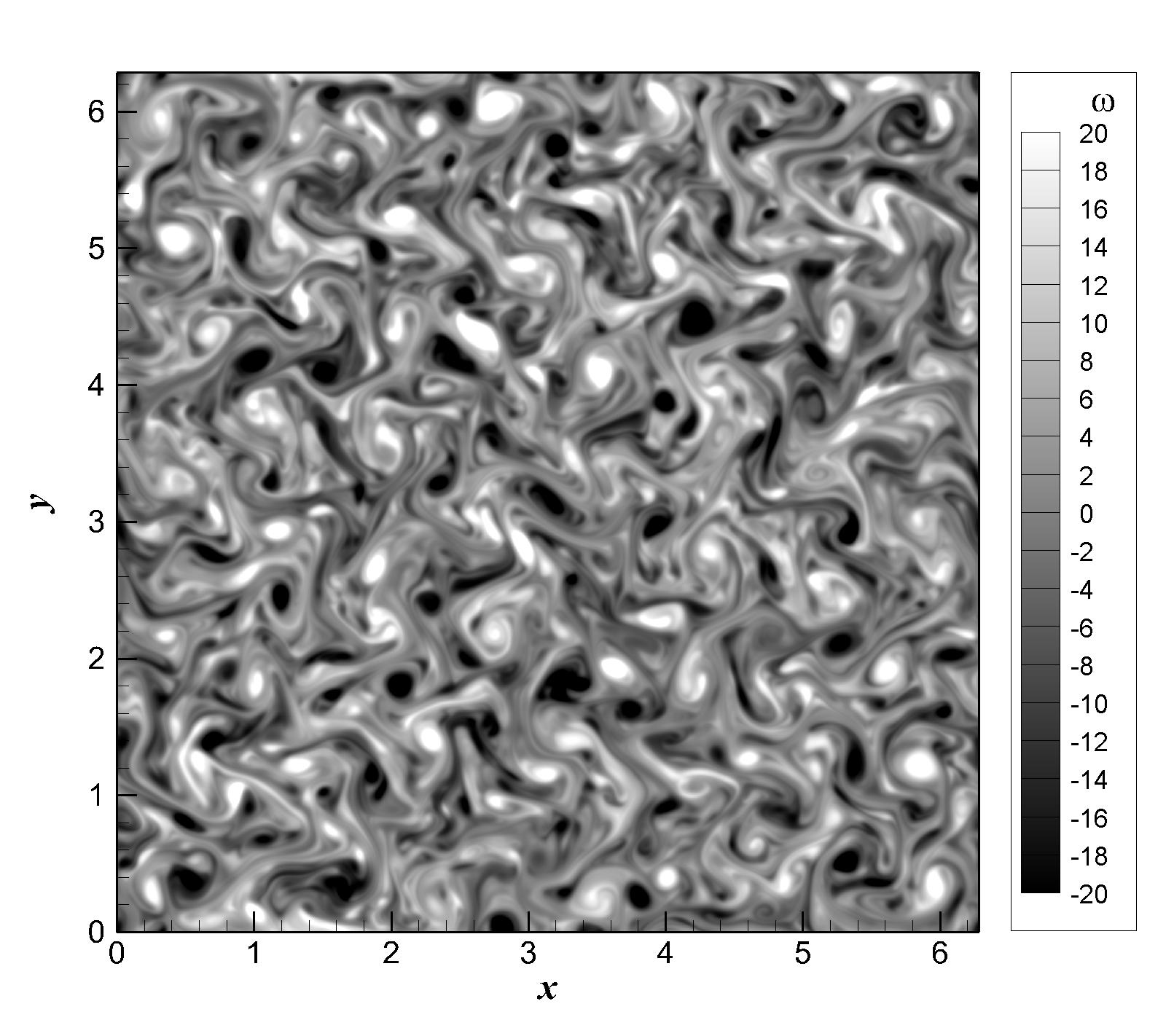}}
\subfigure[$N=5$]{\includegraphics[width=0.5\textwidth]{f_full_df_NA5_r4.jpeg}}
}\\
\mbox{
\subfigure[$N=20$]{\includegraphics[width=0.5\textwidth]{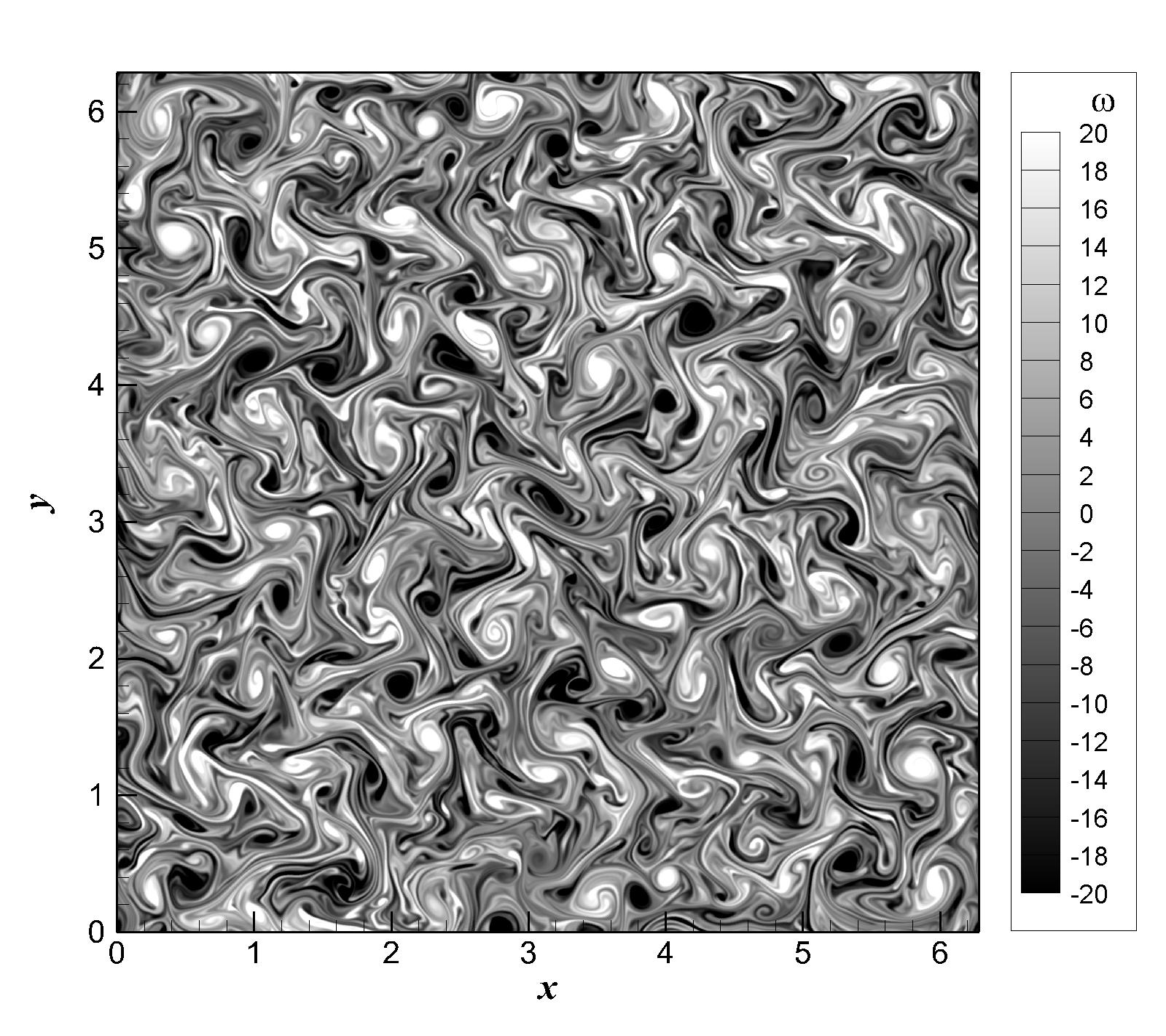}}
\subfigure[$N=100$]{\includegraphics[width=0.5\textwidth]{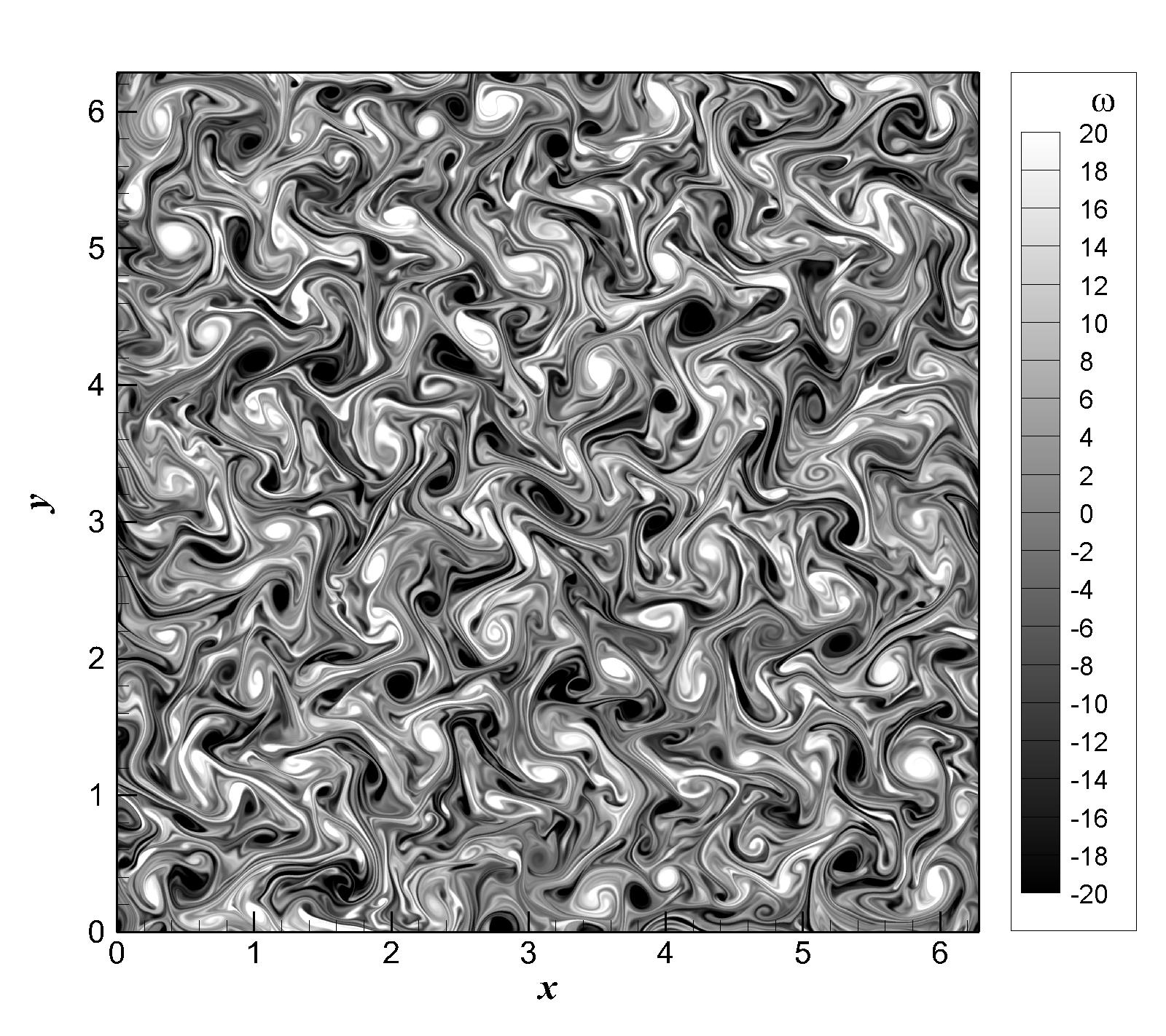}}
}
\caption{Vorticity contour fields for the two-dimensional turbulence data at $Re=32,000$ to illustrate a-priori recovery process of the true data from the filtered data using the BiCGSTAB method by varying the number of iterations $N$.}
\label{f:2dfieldBi}
\end{figure}

\begin{figure}[!ht]
\centering
\mbox{
\subfigure{\includegraphics[width=0.5\textwidth]{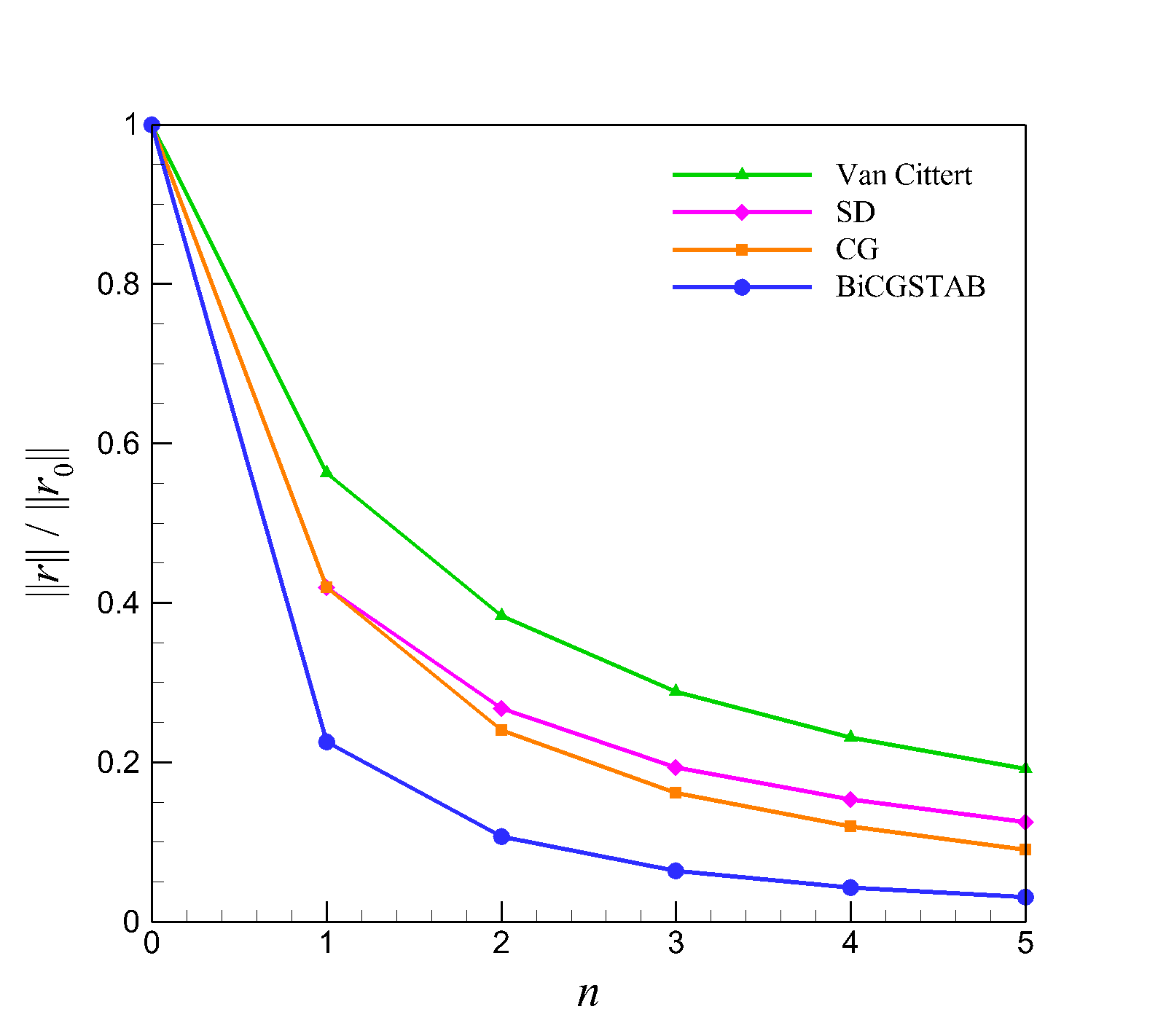}}
\subfigure{\includegraphics[width=0.5\textwidth]{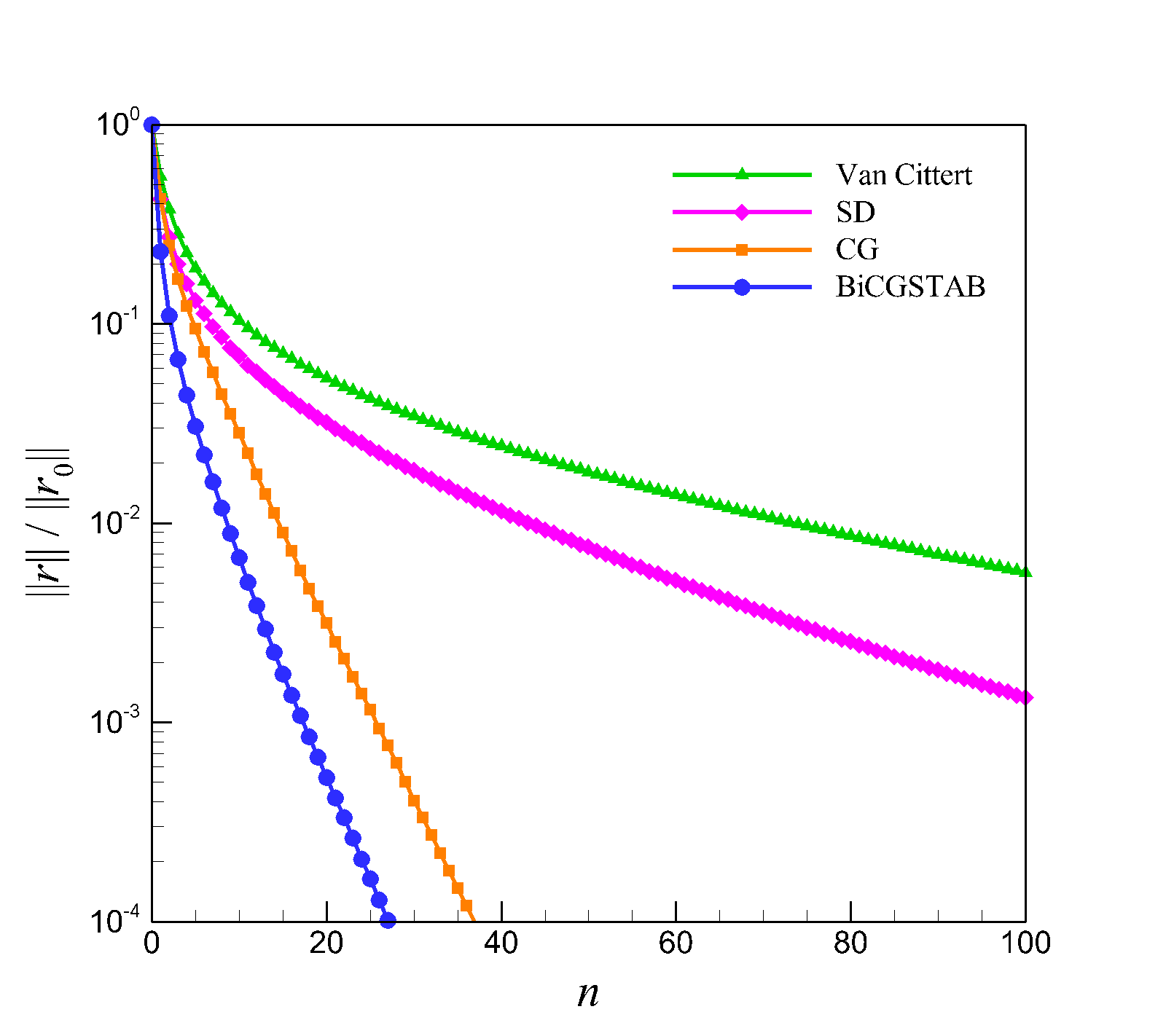}}
}
\caption{Residual history of the iterative approximate deconvolution process for the two-dimensional turbulence data of $2048^2$ (left) and $256^2$ (right) resolutions.}
\label{f:2dres}
\end{figure}

\begin{figure}[!ht]
\centering
\mbox{
\subfigure[$N=5$ ($2048^2$)]{\includegraphics[width=0.5\textwidth]{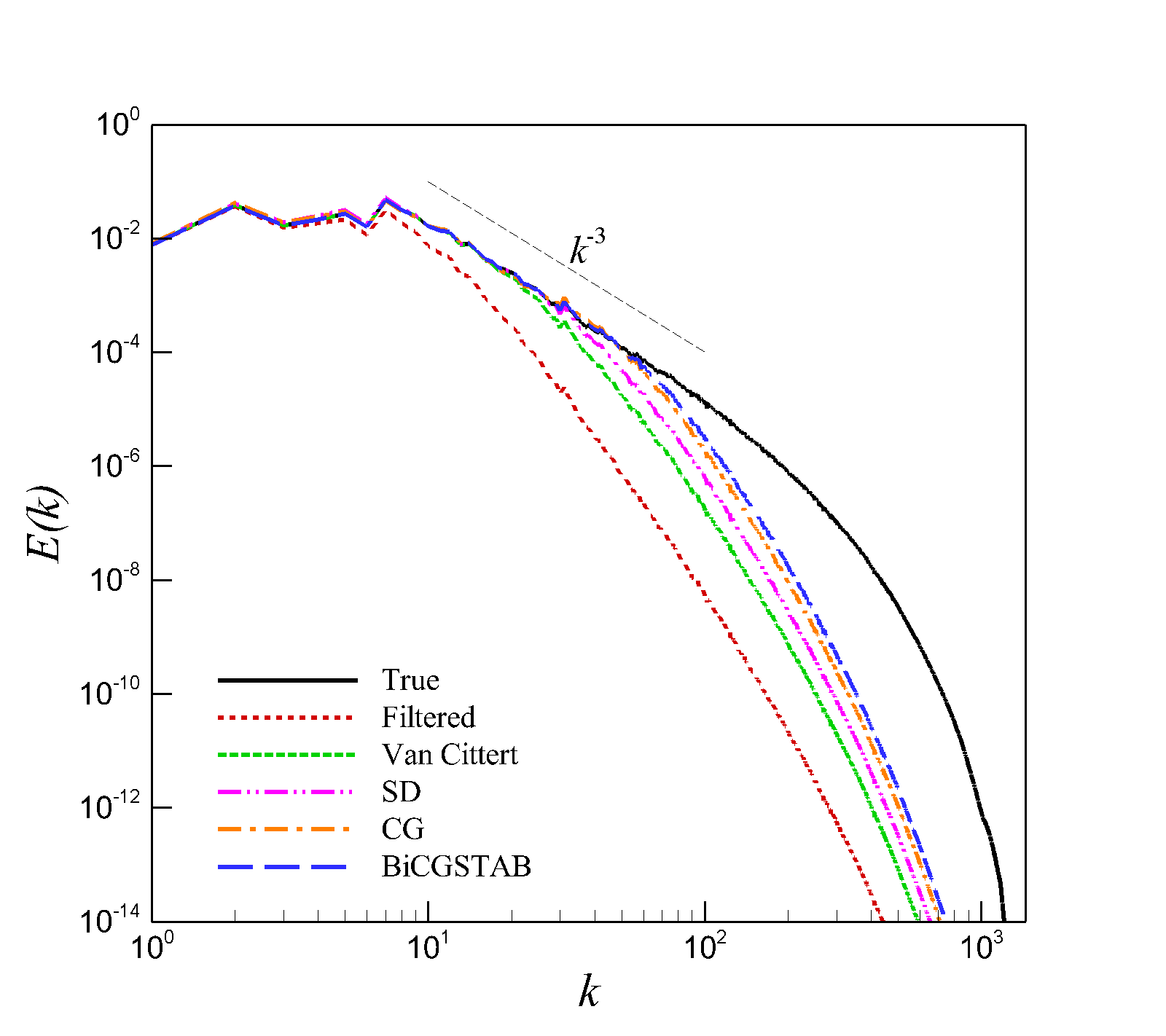}}
\subfigure[BiCGSTAB ($2048^2$)]{\includegraphics[width=0.5\textwidth]{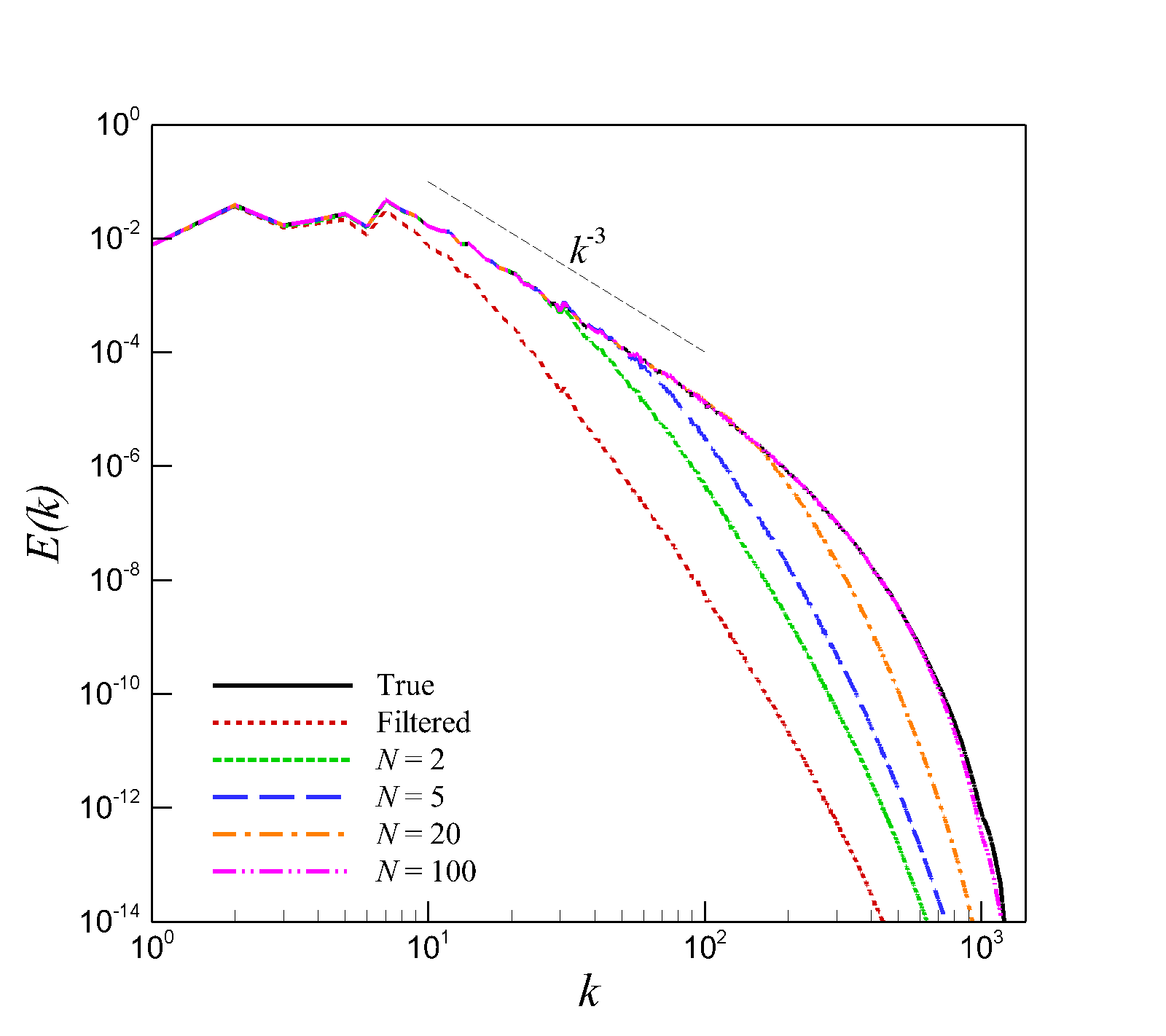}}
}\\
\mbox{
\subfigure[$N=5$ ($256^2$)]{\includegraphics[width=0.5\textwidth]{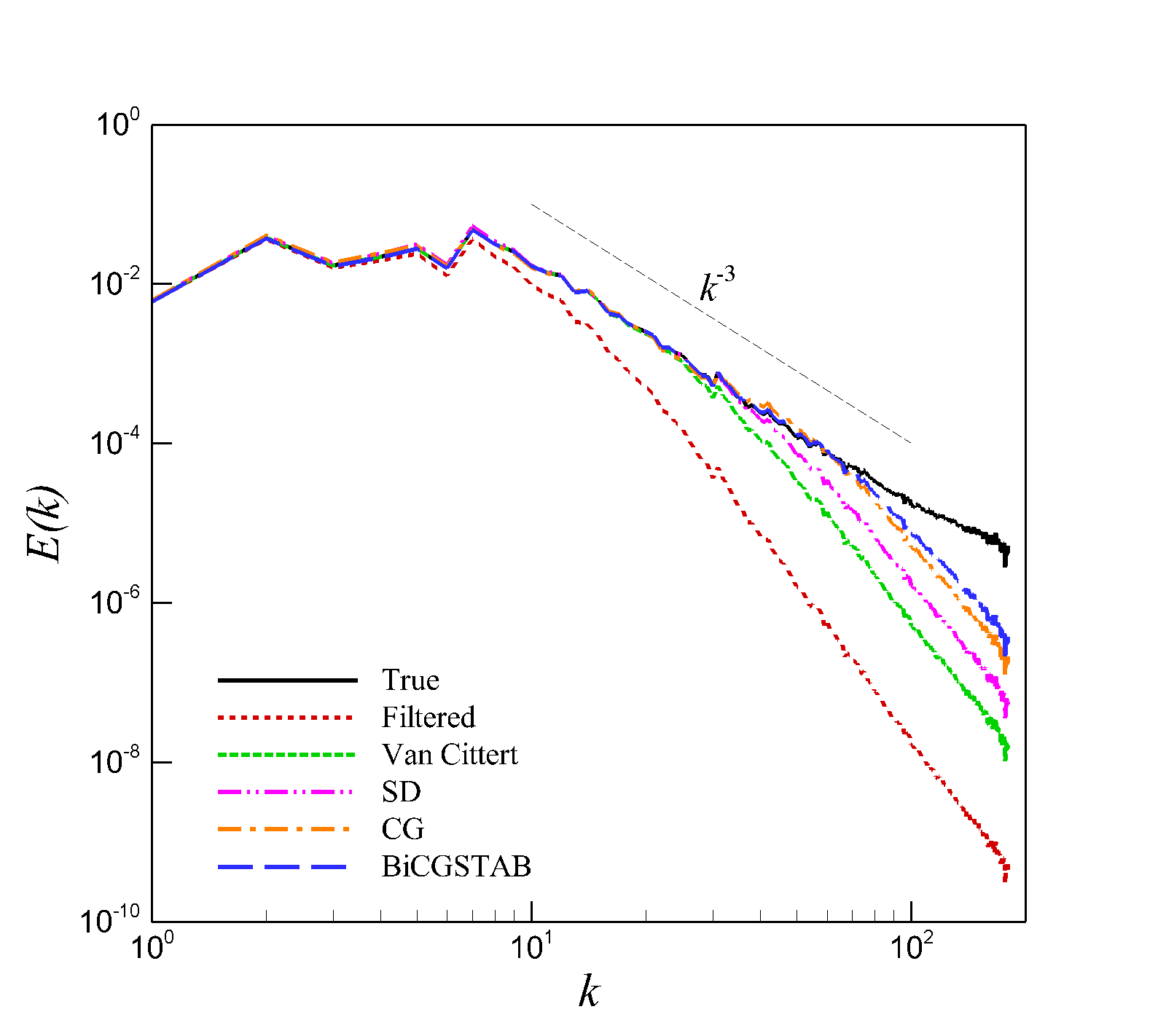}}
\subfigure[BiCGSTAB ($256^2$)]{\includegraphics[width=0.5\textwidth]{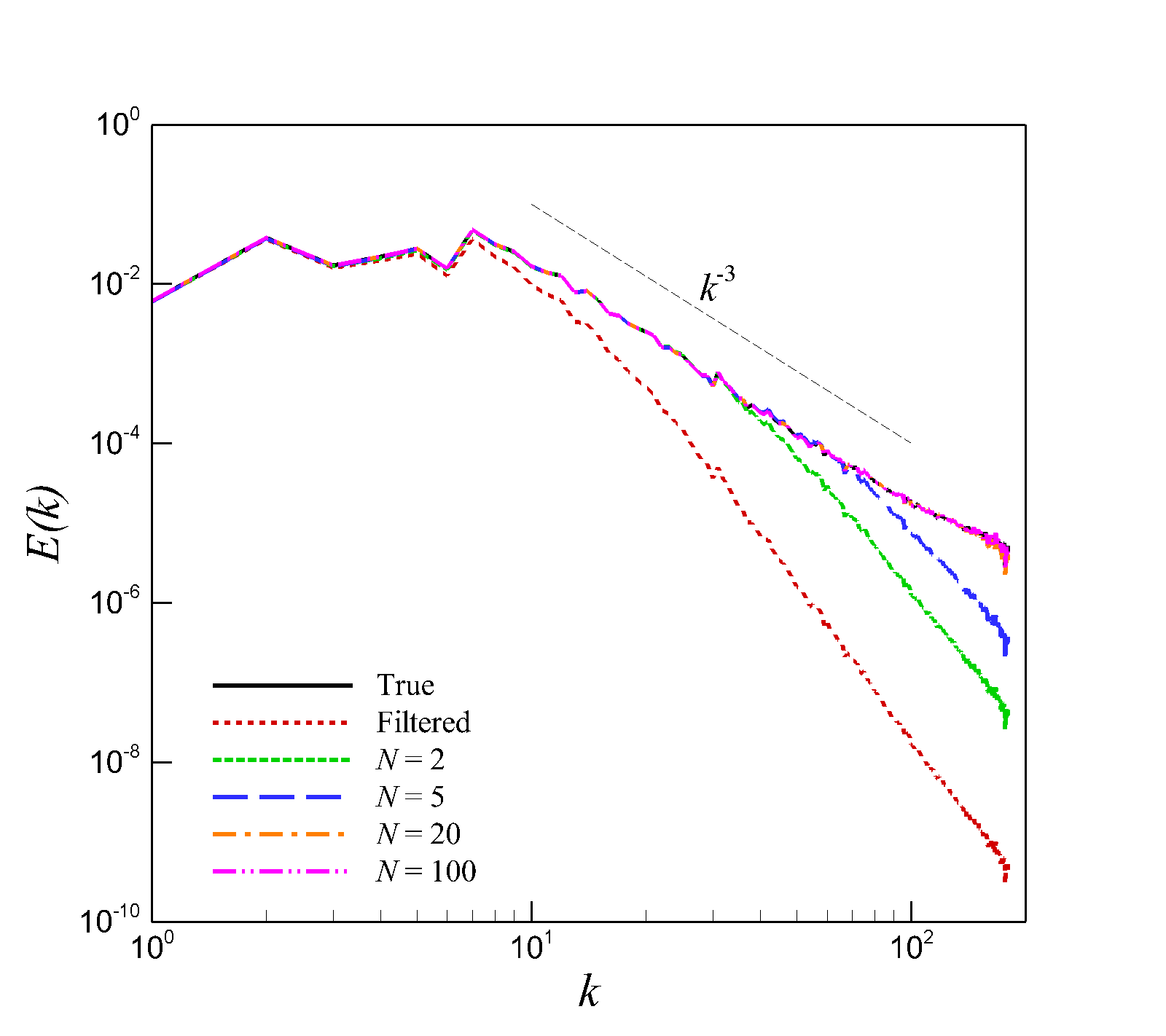}}
}
\caption{Energy spectra of the two-dimensional turbulence data to illustrate recovery performance of the proposed iterative approaches starting from the filtered input data to predict the true data.}
\label{f:2dspec}
\end{figure}

\begin{figure}[!ht]
\centering
\mbox{
\subfigure[$N=5$ ($2048^2$)]{\includegraphics[width=0.5\textwidth]{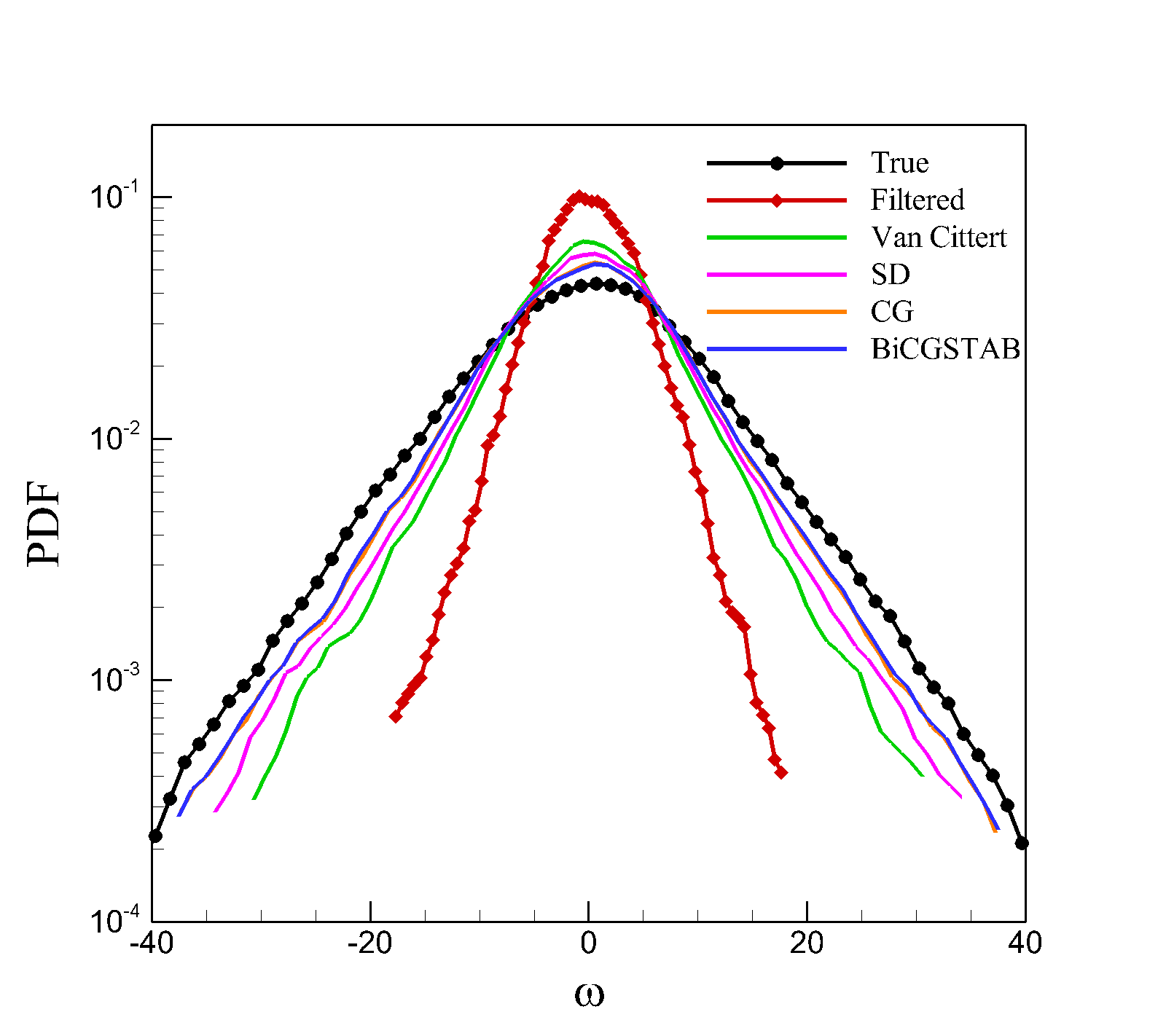}}
\subfigure[BiCGSTAB ($2048^2$)]{\includegraphics[width=0.5\textwidth]{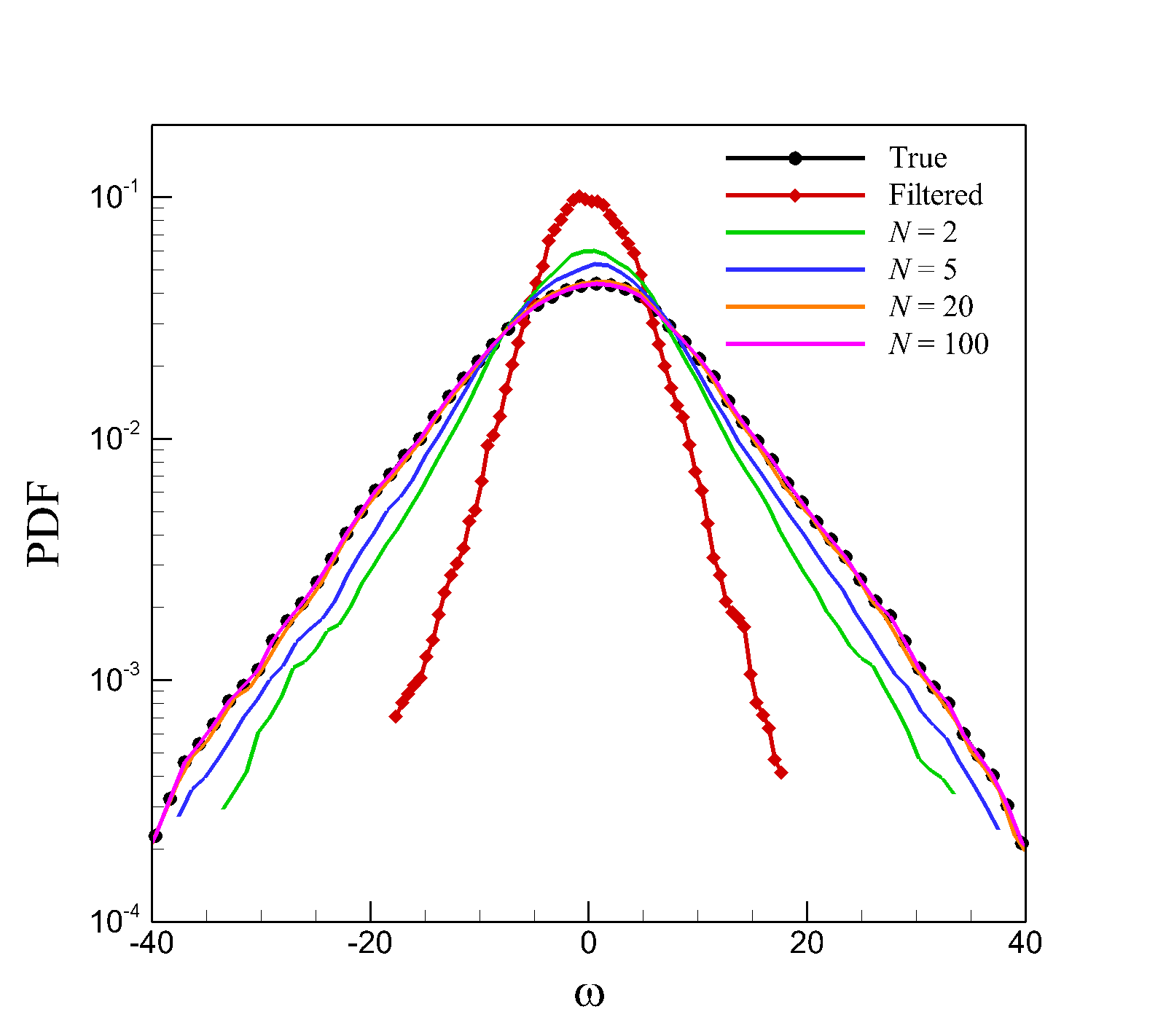}}
}\\
\mbox{
\subfigure[$N=5$ ($256^2$)]{\includegraphics[width=0.5\textwidth]{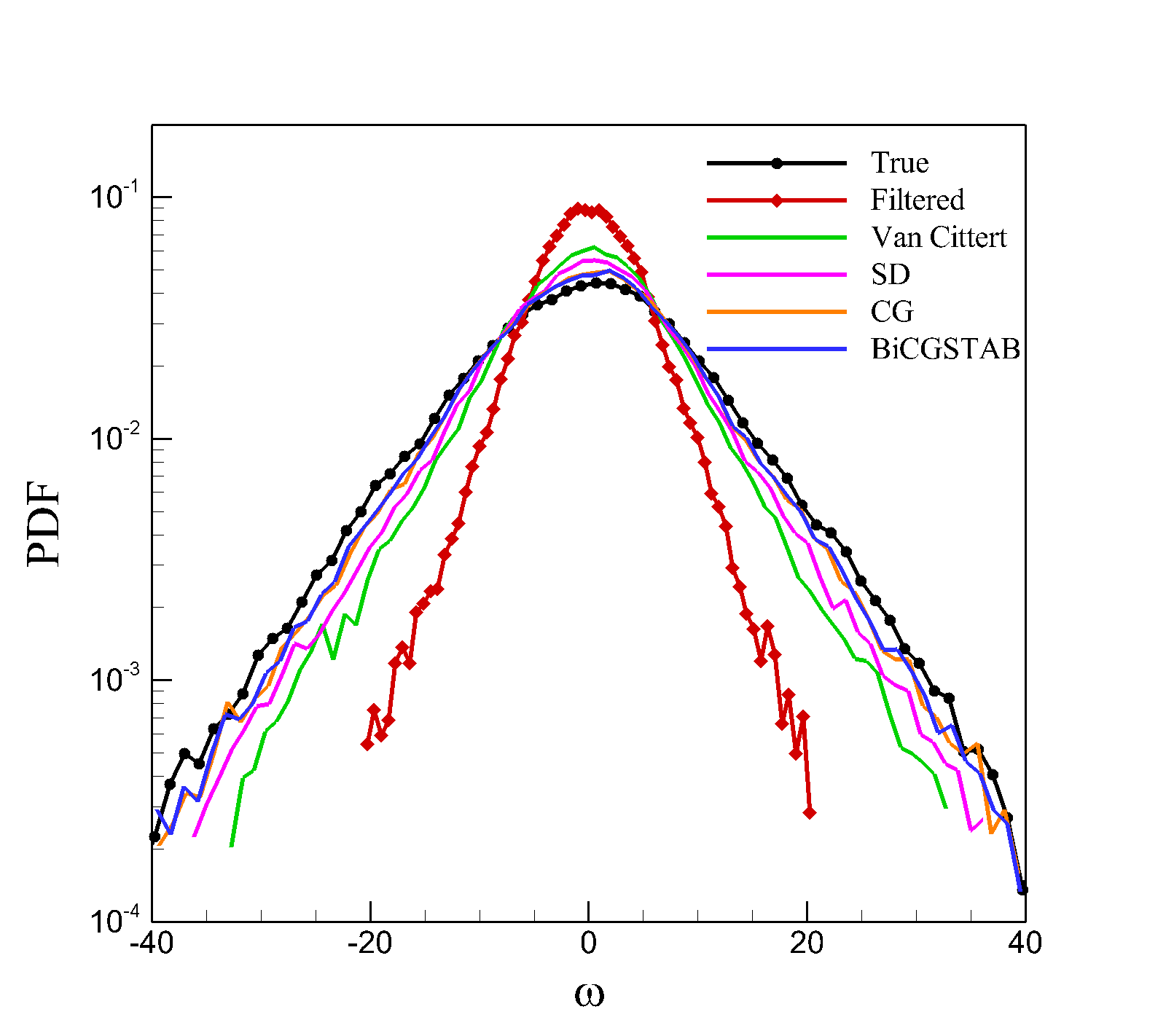}}
\subfigure[BiCGSTAB ($256^2$)]{\includegraphics[width=0.5\textwidth]{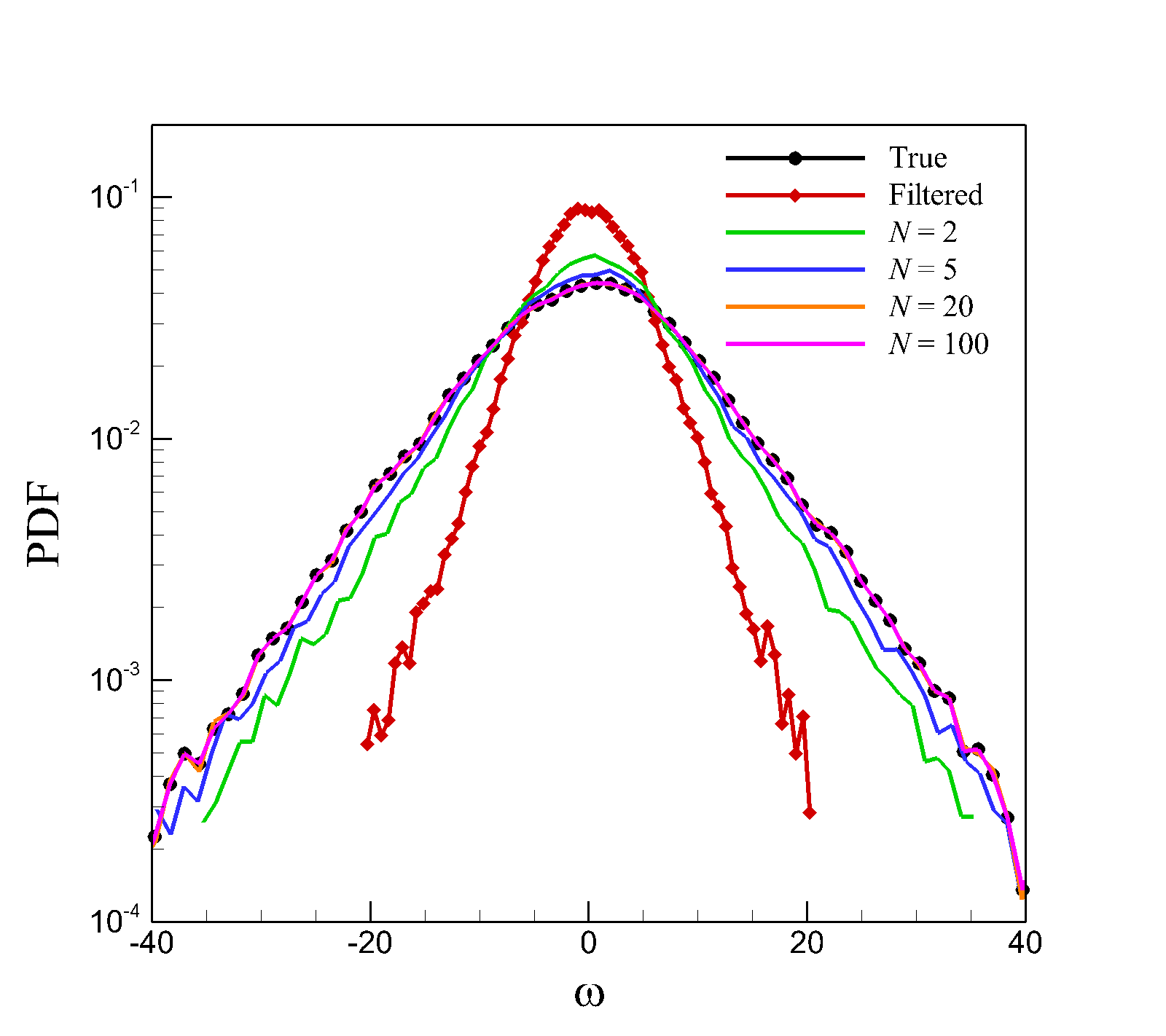}}
}
\caption{Probability density function (PDF) of the two-dimensional turbulence data to illustrate recovery performance of the proposed iterative approaches starting from the filtered input data to predict the true data.}
\label{f:2dpdf}
\end{figure}


Fig.~\ref{f:2dfield5} demonstrates the performance of the proposed algorithms on the fine resolution space. The true data is filtered by using the Germano's elliptic differential filter with $\gamma=0.01$, and recovered results by all four algorithms presented in Section~\ref{s:gad} are shown for $N=5$. We highlight that all the proposed deconvolution algorithms start using the same filtered field data as an initial condition for the iterative process. It can easily be seen that the CG and BiCGSTAB methods predict significantly improved results; almost all fine scale details are recovered back. Furthermore,  Fig.~\ref{f:2dfieldBi} presents the vorticity fields for the same two-dimensional turbulence data at $Re=32,000$ to illustrate a-priori recovery process of the true data from the filtered data using the BiCGSTAB algorithm with varying the number of iteration $N$. This sensitivity analysis demonstrates that the iterative procedure successfully converges to the true solution. Furthermore, a normalized (i.e., with respect to the initial residual) residual history comparison of the iterative processes is plotted in Fig.~\ref{f:2dres} for both fine and coarse resolutions. It can be easily seen that the Van Cittert and the steepest descent method yield substantially slow convergence rate when compared to the conjugate gradient algorithms. It is clear that even the first iteration of the BiCGSTAB algorithm provides more accurate recovered data compared to the Van Cittert process after $N=5$ iterations.

Fig.~\ref{f:2dspec} shows the performance of the proposed methodology in terms of statistical assessments given by averaged kinetic energy spectra and probability density functions (PDF) of the vorticity field. Here, we present our results for both high and coarse resolution data sets. The filtering parameter that controls the filtering radius are set $\gamma=0.01$ and $\gamma=0.1$ for the high-fidelity and coarse-grained data sets, respectively. Our analysis includes a comparison of all models for $N=5$ and a sensitivity analysis with respect to $N$ when using the BiCGSTAB iterative model. For sub-filter scale reconstruction performance assessment, it can readily be observed that the conjugate gradient based methods manage to recover a far greater region of the inertial range in accordance with the $k^{-3}$ scaling law. This may also be observed from the PDF comparisons where the narrow band distribution caused by the low-pass spatial filtering is successfully recovered to its true spread.

\begin{table}[!t]
\centering
\caption{Computational efficiency of the proposed deconvolution methods. Compared with the standard Van Cittert method, note that both the CG and BiCGSTAB methods provide approximately four times faster results for 99\% reduction of the residual.}
\label{tab:1}
\begin{tabular}{lrrcc}
\hline\noalign{\smallskip}
\multicolumn{1}{l}{AD Model} &
\multicolumn{2}{c}{\underline{~~~~~~~~CPU time/iteration~~~~~~~~~}}&
\multicolumn{2}{c}{\underline{CPU time (99\% recovery)}} \\ \noalign{\smallskip}
 &   \multicolumn{1}{c}{2D} & \multicolumn{1}{c}{3D} & \multicolumn{1}{c}{2D} & \multicolumn{1}{c}{3D}  \\ \noalign{\smallskip} \hline\noalign{\smallskip}
Van Cittert             & 1.5156$\times10^{-2}$  &  6.9661$\times10^{-2}$   & 1.1064  & 6.6178    \\
Steepest Descent (SD)   & 1.5625$\times10^{-2}$  &  7.6068$\times10^{-2}$   & 0.6719  & 3.9555   \\
Conjugate Gradient (CG) & 1.6093$\times10^{-2}$  &  8.3464$\times10^{-2}$   & 0.2414  & 1.2520    \\
BiCGSTAB                & 3.0625$\times10^{-2}$  &  16.3724$\times10^{-2}$  & 0.2450  & 1.3098   \\
\noalign{\smallskip}\hline
\end{tabular}
\end{table}

Next we present our measurements for computational efficiency of the proposed algorithms. Table~\ref{tab:1} documents CPU time per each iteration for these four algorithms. It can be seen that computational cost for the Van Cittert, steepest descent and conjugate gradient algorithms are similar. Compared to the Van Cittert algorithm, only fractions of 3\% and 6\% overhead estimates are observed by the steepest descent and conjugate gradient algorithms, respectively. This is due to the fact that these algorithms only require one call for the filtering operator per each iteration. However, the computational time required for the BiCGSTAB algorithm becomes double since it requires two calls for the filtering operator in each iteration. Although we observe higher computational cost for the BiCGSTAB algorithm per each subsequent iteration, as shown in Table~\ref{tab:1}, a prediction of 99\% data recovery can be efficiently obtained by using the BiCGSTAB method due to its superior convergence rate (i.e., please see Fig.~\ref{f:2dres} as well). Therefore, we conclude that both the standard conjugate gradient (CG) and stabilized biconjugate gradient (BiCGSTAB) methods provide similar performance characteristics which are significantly superior than the Van Cittert and steepest descent methods. This offers a great promise of using conjugate gradient based methods for the approximate deconvolution structural models.

\subsection{Three-dimensional Kolmogorov turbulence}
\label{s:3D}

\begin{figure}[!ht]
\centering
\mbox{
\subfigure[True]{\includegraphics[width=0.5\textwidth]{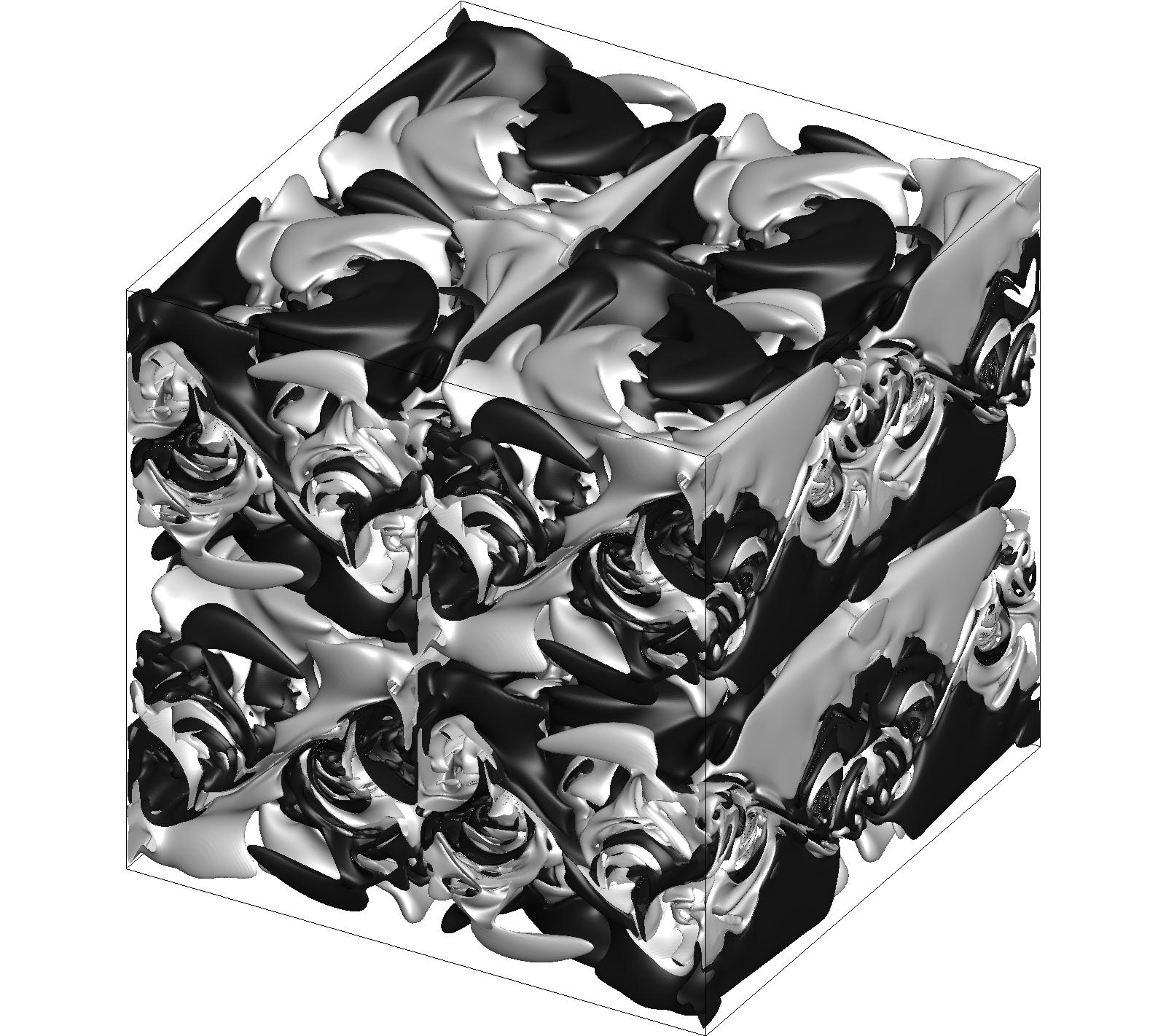}}
\subfigure[Filtered]{\includegraphics[width=0.5\textwidth]{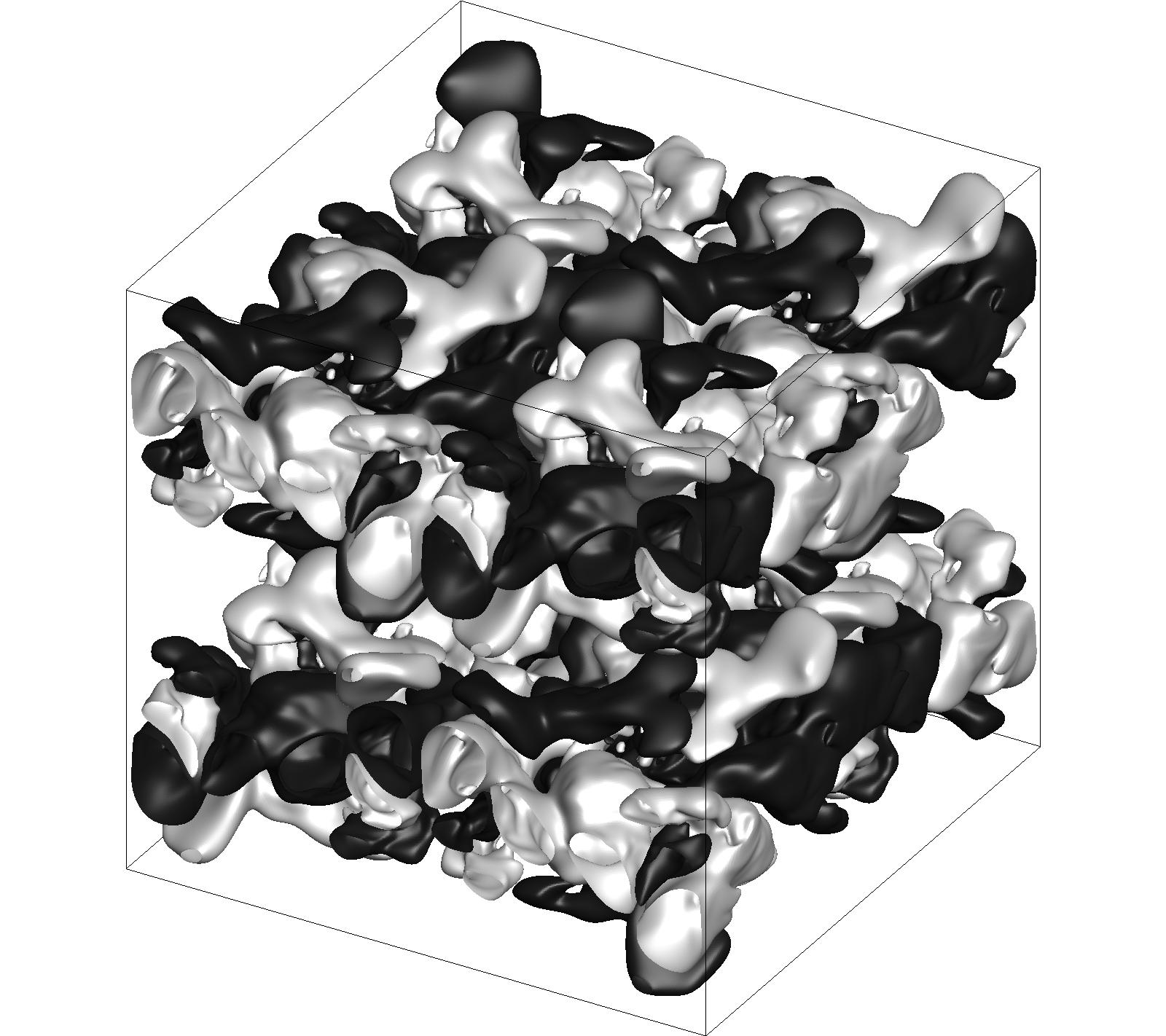}}
}\\
\mbox{
\subfigure[Van Cittert]{\includegraphics[width=0.5\textwidth]{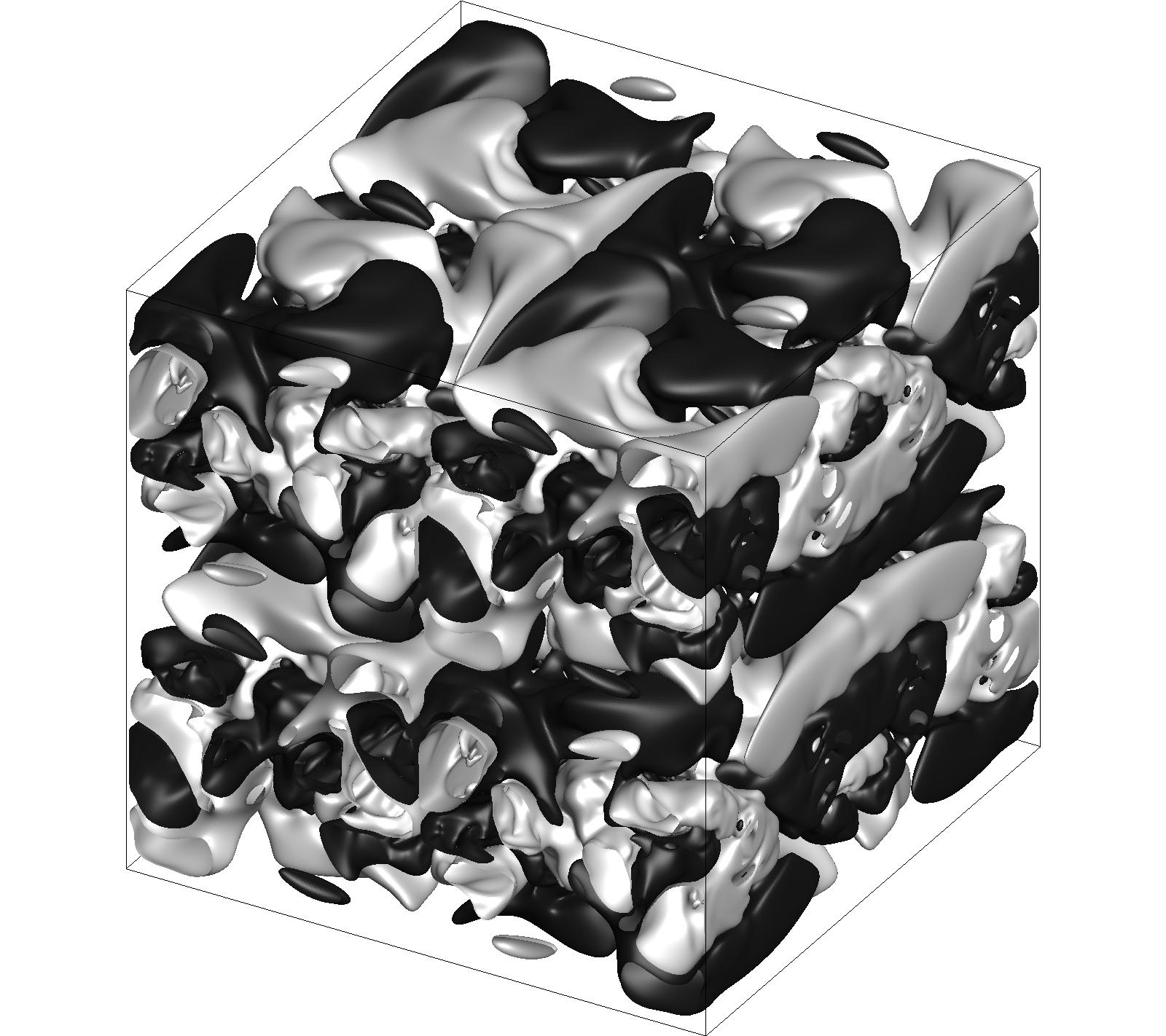}}
\subfigure[Steepest Descent]{\includegraphics[width=0.5\textwidth]{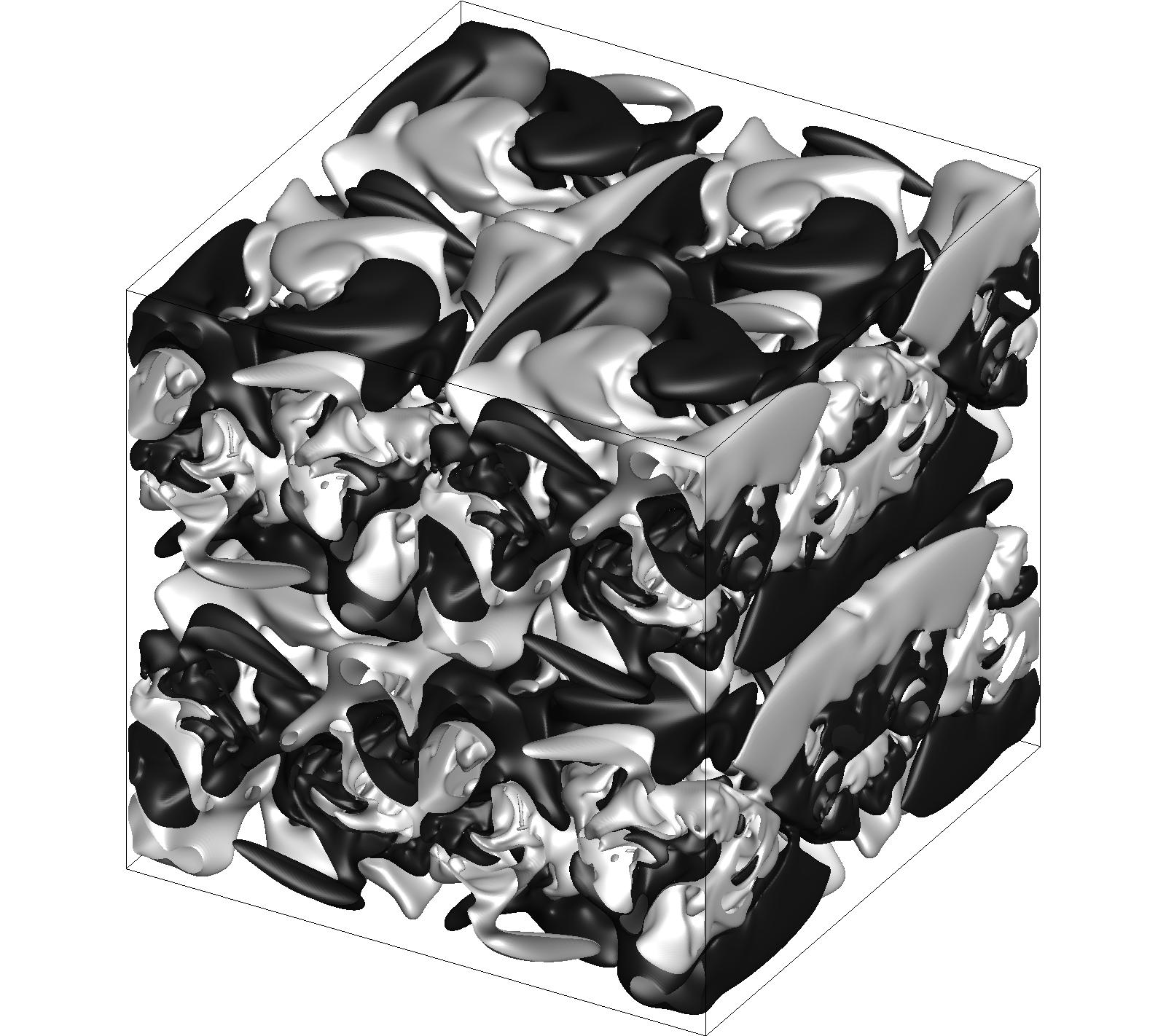}}
}\\
\mbox{
\subfigure[Conjugate Gradient]{\includegraphics[width=0.5\textwidth]{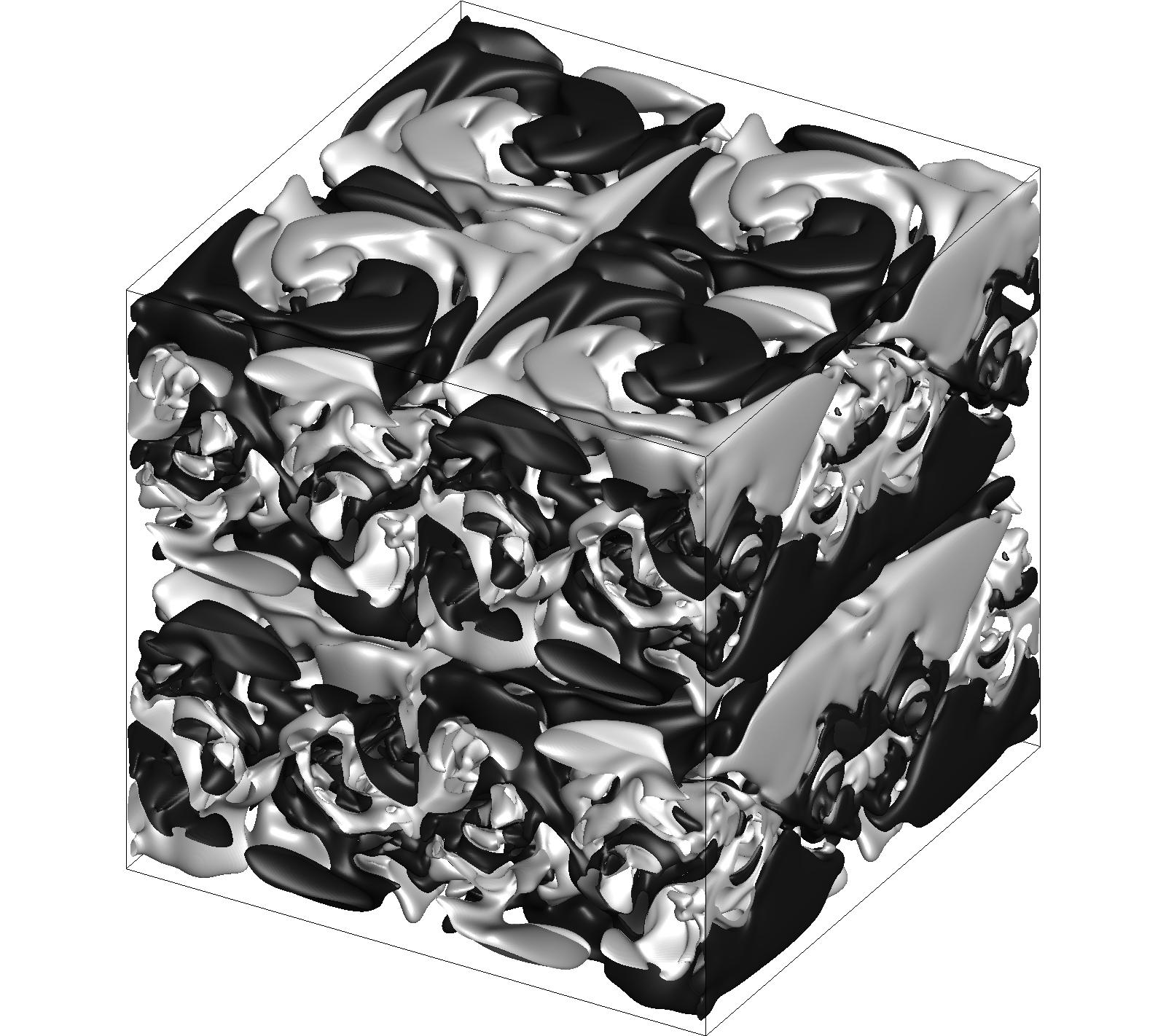}}
\subfigure[BiCGSTAB]{\includegraphics[width=0.5\textwidth]{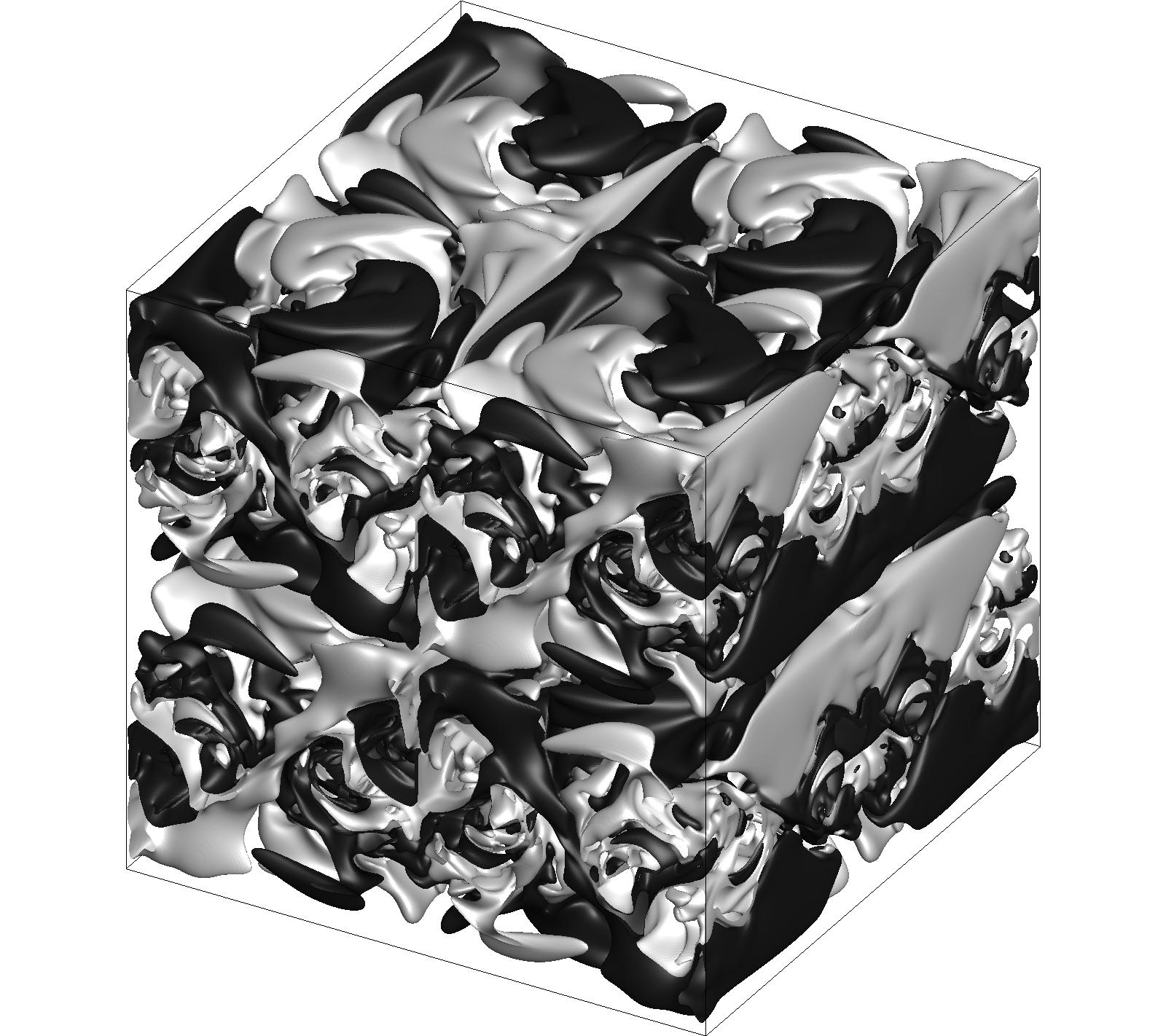}}
}
\caption{Isosurfaces of $\omega_x = \pm 0.3$ for the three-dimensional turbulence data at $Re=1600$ to illustrate a-priori recovery process of the true data from the filtered data using $N=5$ iterations.}
\label{f:3disoN}
\end{figure}

\begin{figure}[!ht]
\centering
\mbox{
\subfigure[True]{\includegraphics[width=0.5\textwidth]{g_w1_true.jpeg}}
\subfigure[Filtered]{\includegraphics[width=0.5\textwidth]{g_w1_full_df.jpeg}}
}\\
\mbox{
\subfigure[$N=2$]{\includegraphics[width=0.5\textwidth]{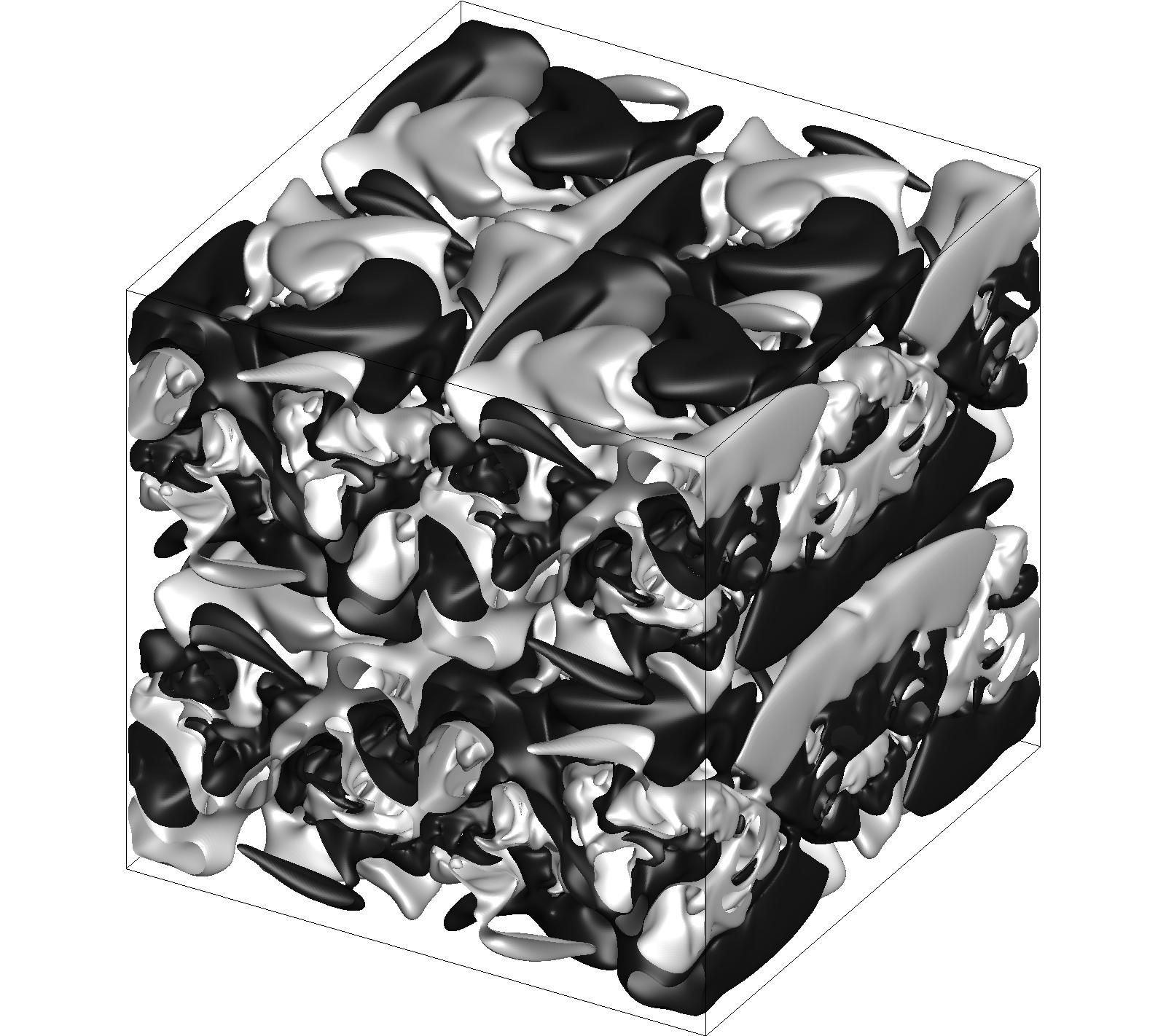}}
\subfigure[$N=5$]{\includegraphics[width=0.5\textwidth]{g_w1_full_df_NA5_r4.jpeg}}
}\\
\mbox{
\subfigure[$N=20$]{\includegraphics[width=0.5\textwidth]{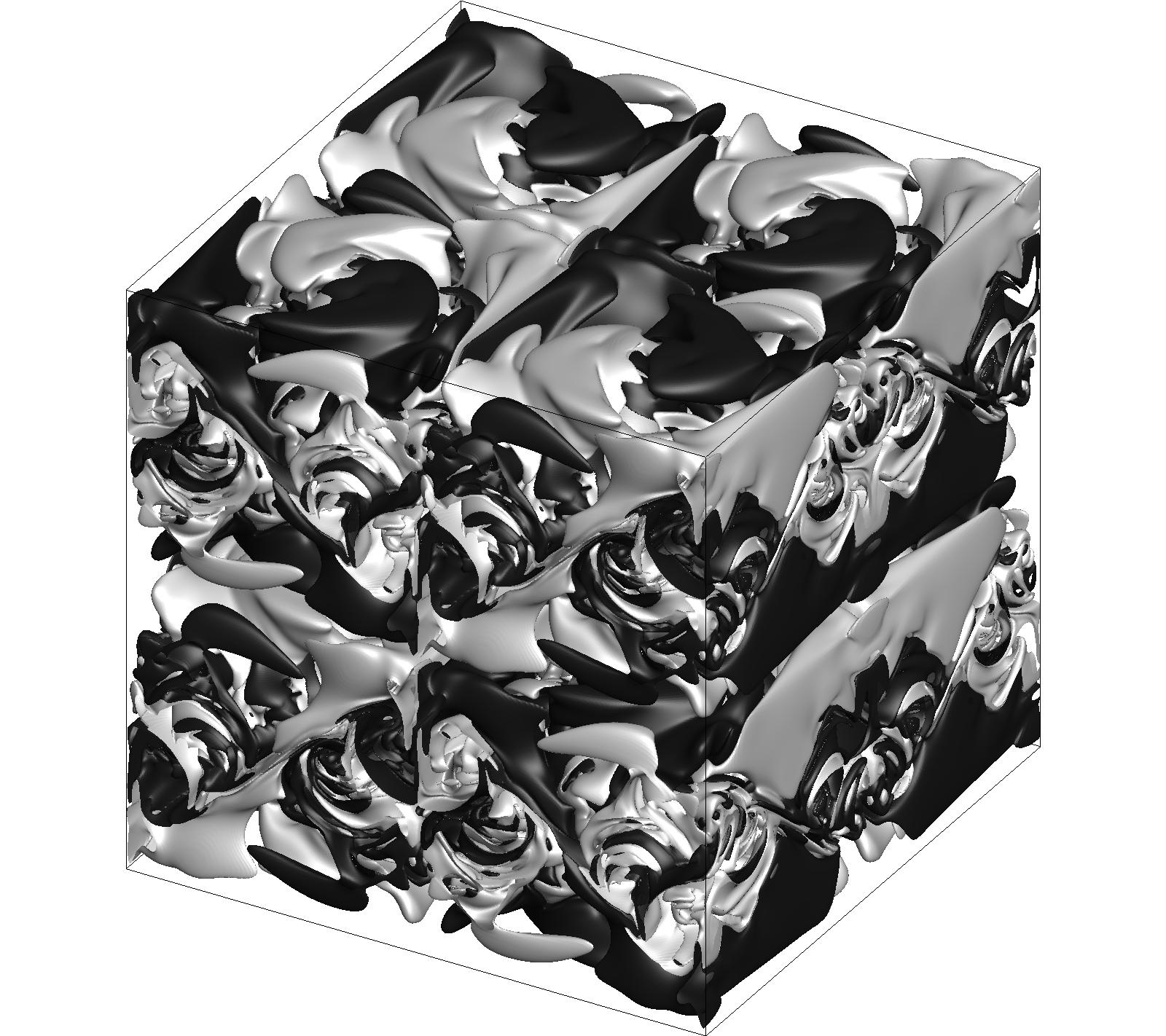}}
\subfigure[$N=100$]{\includegraphics[width=0.5\textwidth]{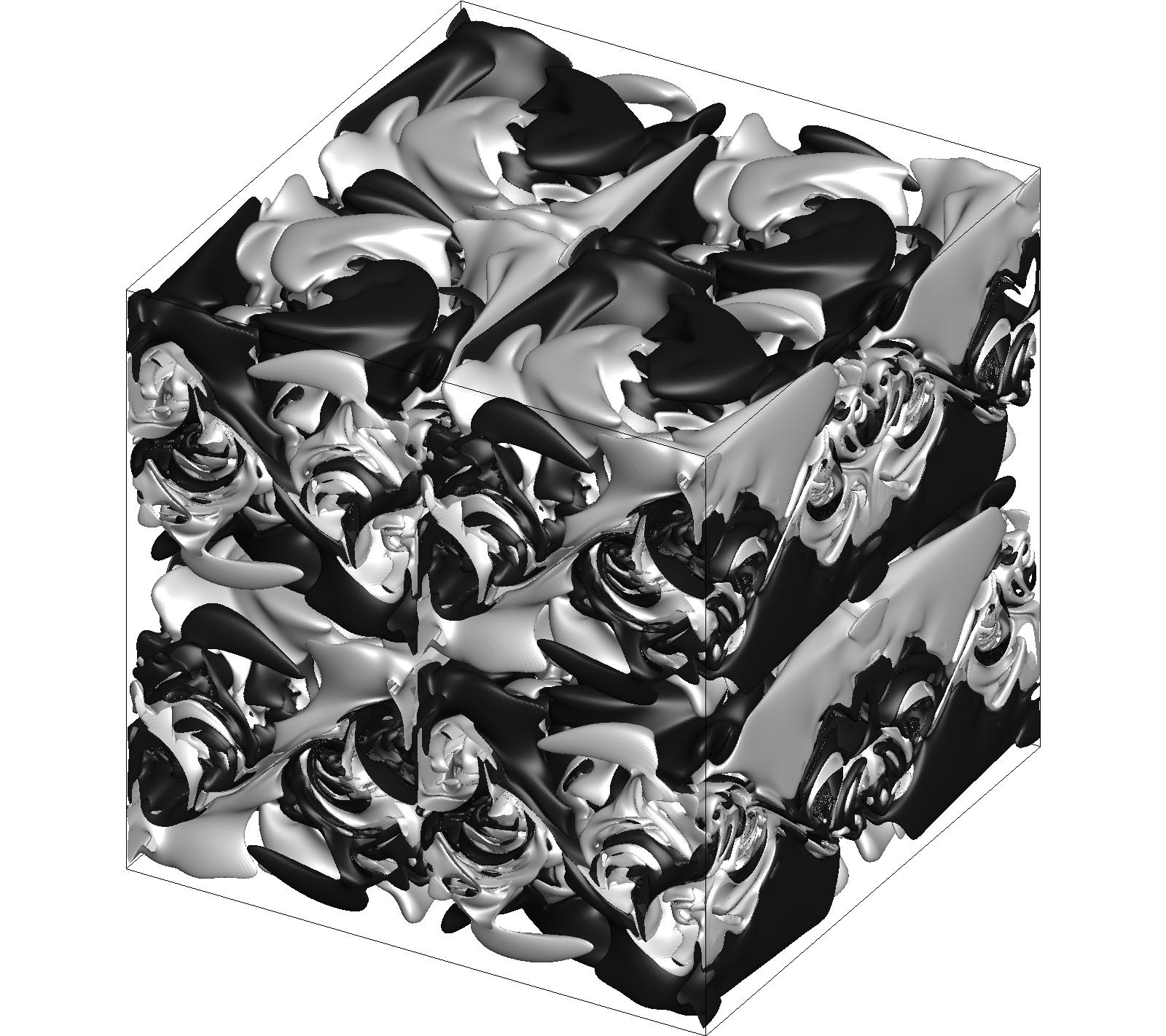}}
}
\caption{Isosurfaces of $\omega_x = \pm 0.3$ for the three-dimensional turbulence data at $Re=1600$ to illustrate a-priori recovery process of the true data from the filtered data using the BiCGSTAB method by varying the number of iterations $N$.}
\label{f:3disoBi}
\end{figure}

\begin{figure}[!ht]
\centering
\mbox{
\subfigure{\includegraphics[width=0.5\textwidth]{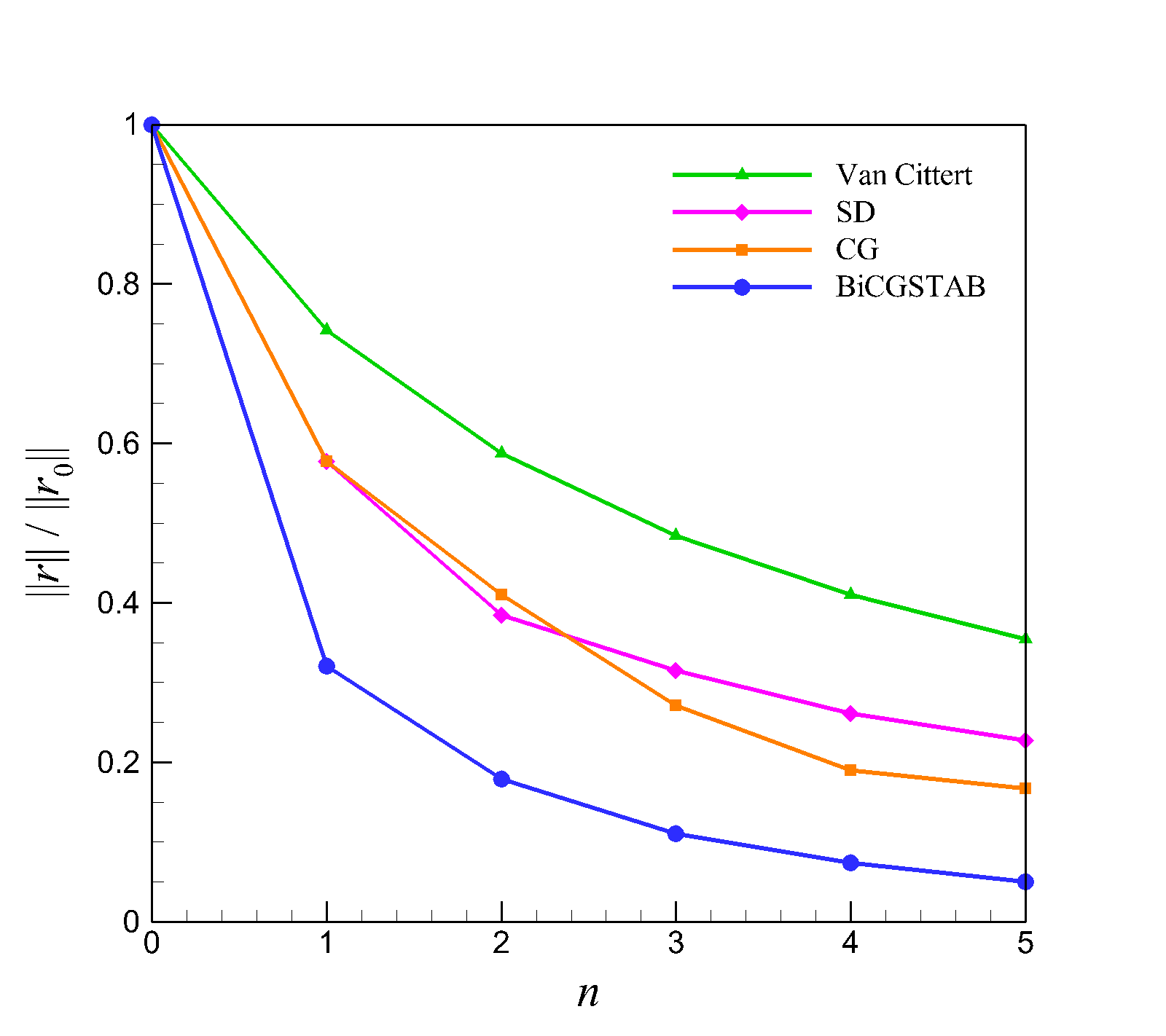}}
\subfigure{\includegraphics[width=0.5\textwidth]{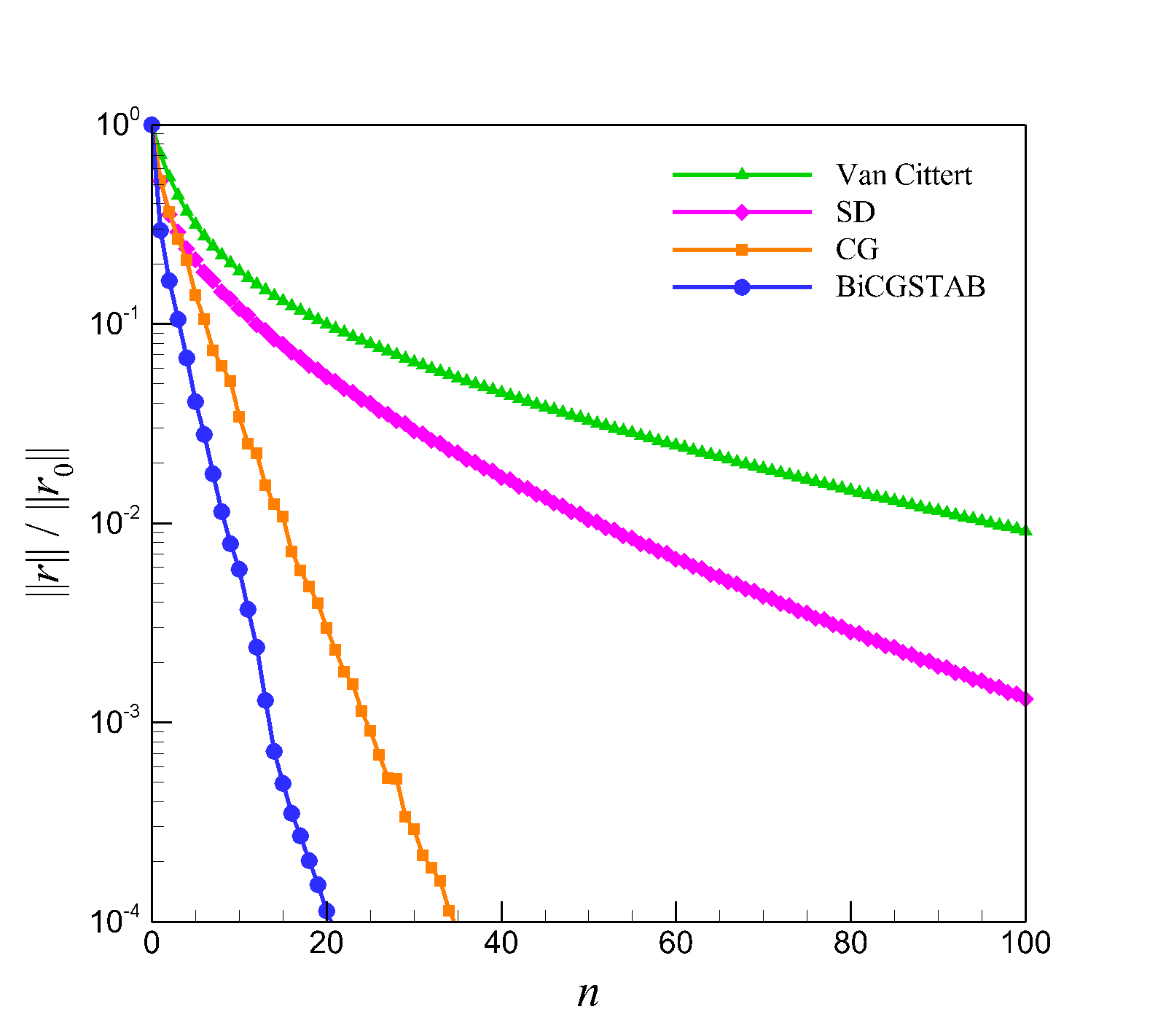}}
}
\caption{Residual history of the iterative approximate deconvolution processes for the three-dimensional turbulence data of $512^3$ (left) and $64^3$ (right) resolutions.}
\label{f:3dres}
\end{figure}

\begin{figure}[!ht]
\centering
\mbox{
\subfigure[$N=5$ ($512^3$)]{\includegraphics[width=0.5\textwidth]{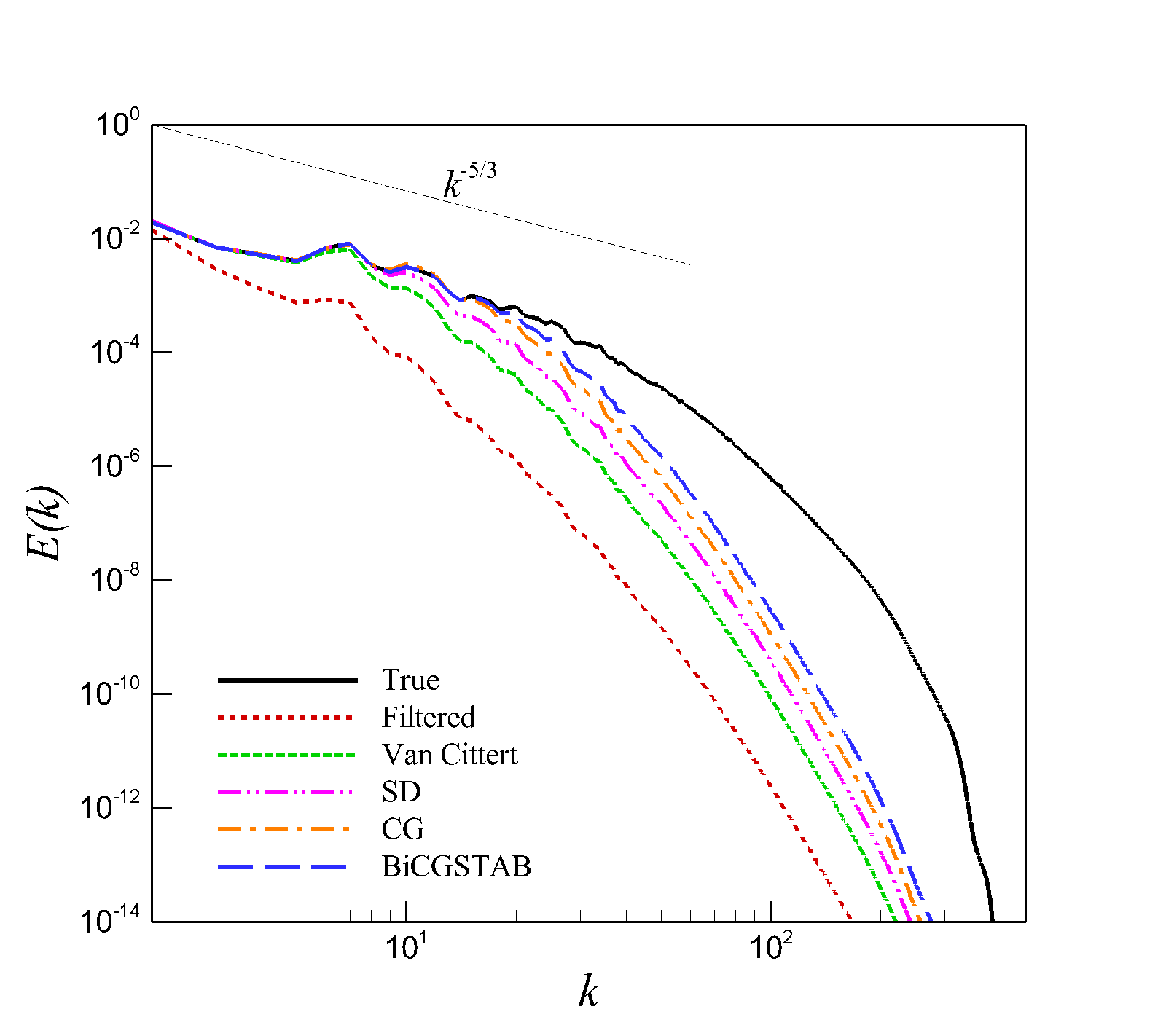}}
\subfigure[BiCGSTAB ($512^3$)]{\includegraphics[width=0.5\textwidth]{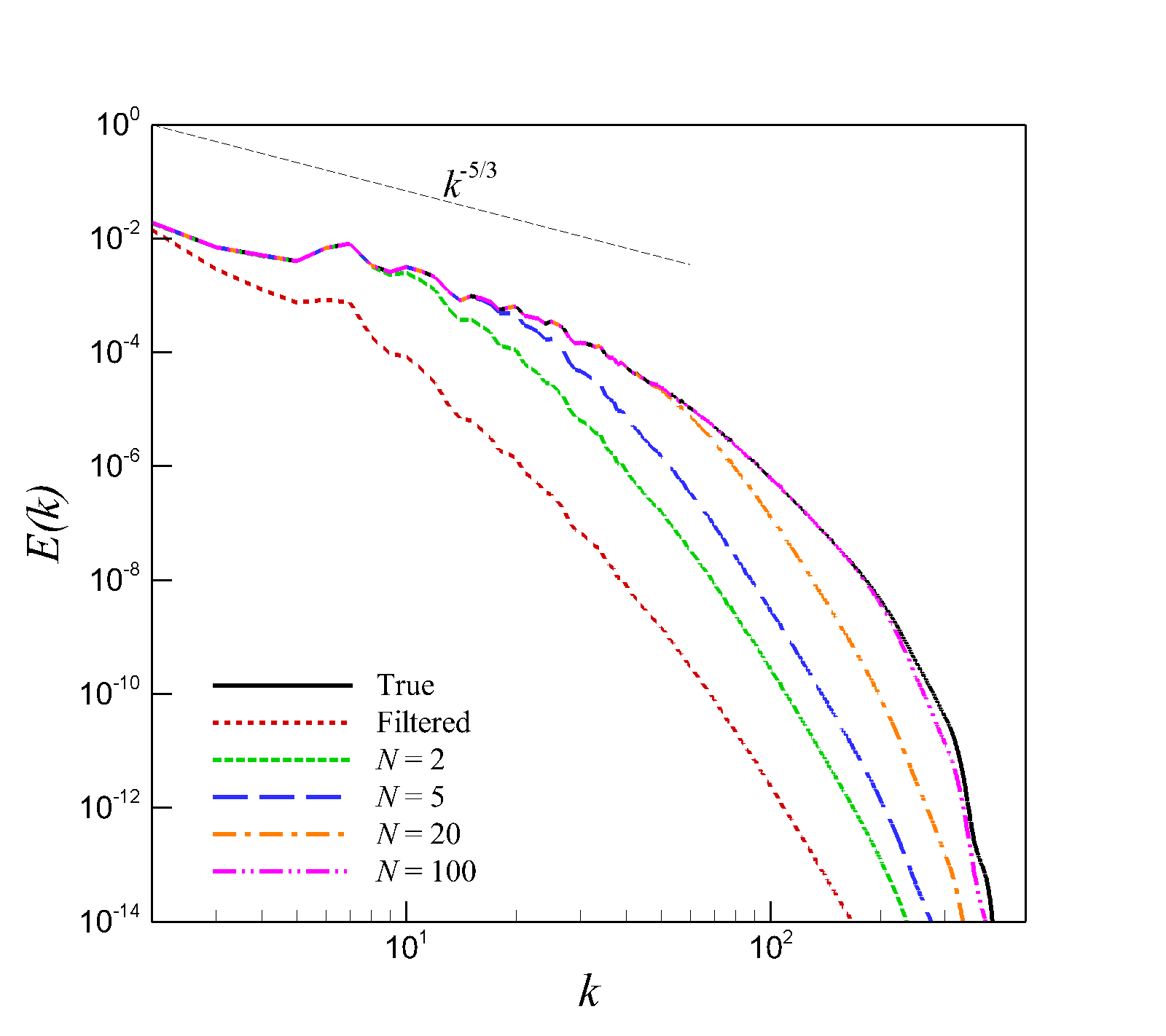}}
}\\
\mbox{
\subfigure[$N=5$ ($64^3$)]{\includegraphics[width=0.5\textwidth]{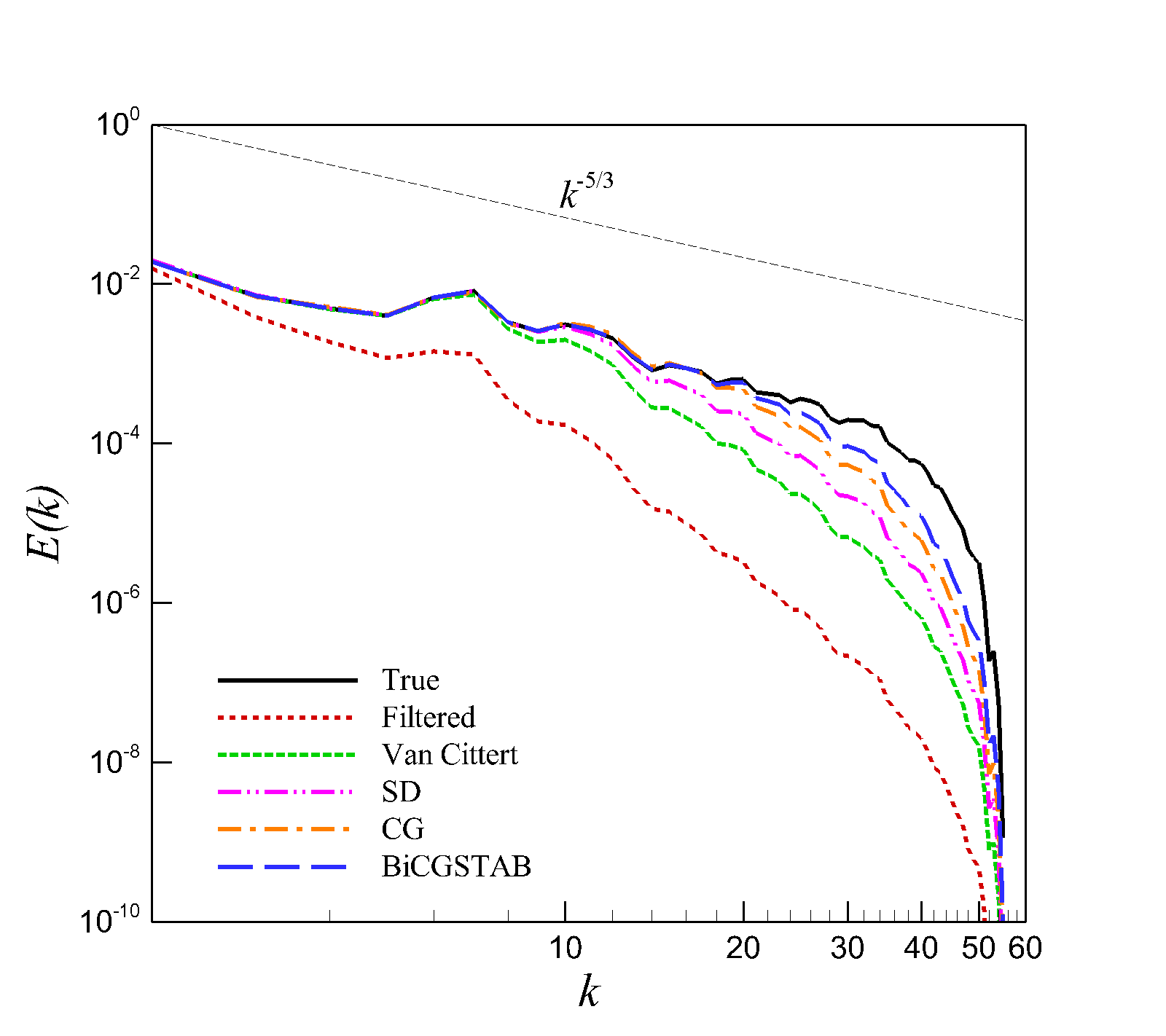}}
\subfigure[BiCGSTAB ($64^3$)]{\includegraphics[width=0.5\textwidth]{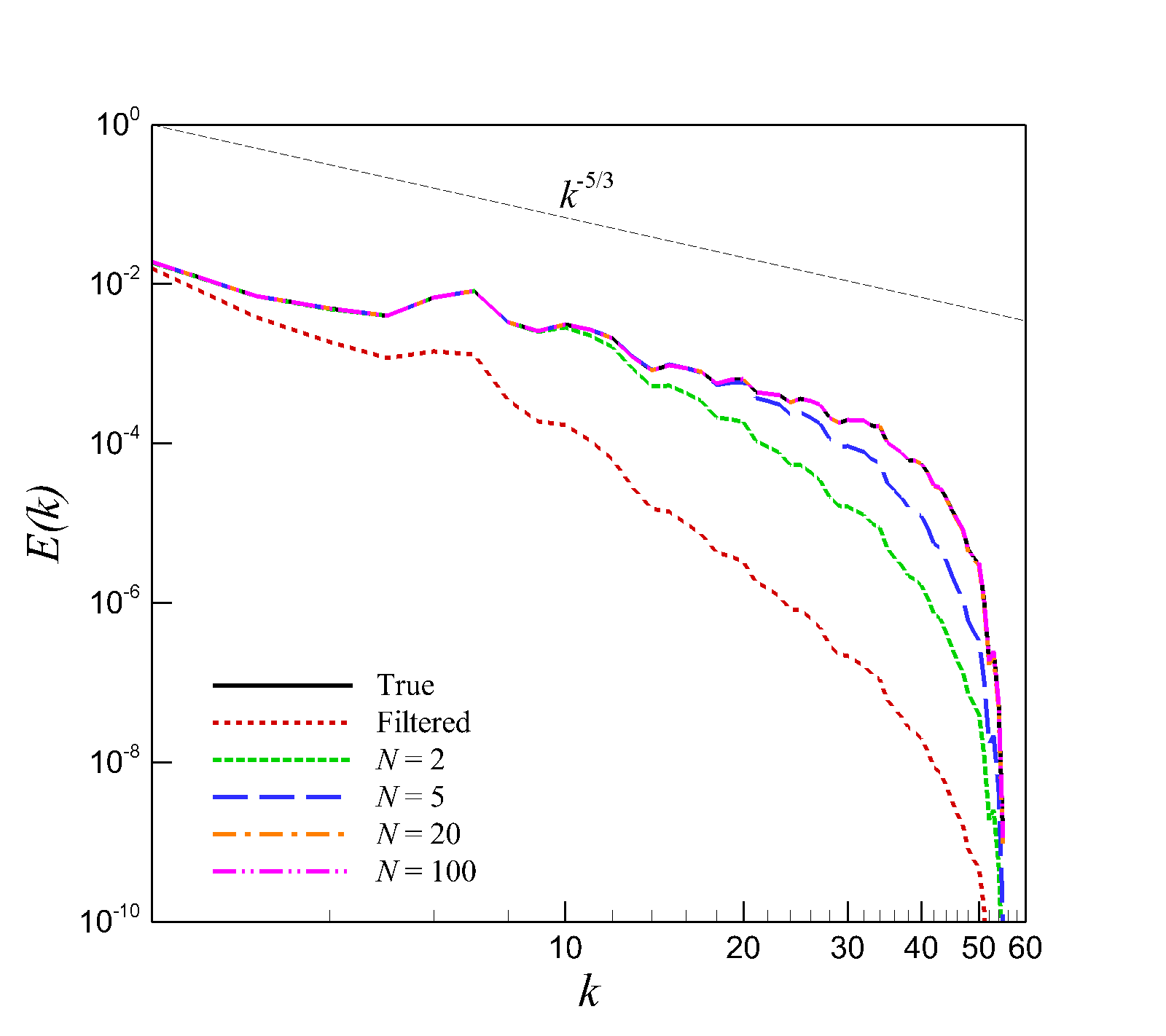}}
}
\caption{Energy spectra of the three-dimensional turbulence data to illustrate recovery performance of the proposed iterative approaches starting from the filtered input data to predict the true data.}
\label{f:3dspec}
\end{figure}

\begin{figure}[!ht]
\centering
\mbox{
\subfigure[$N=5$ ($512^3$)]{\includegraphics[width=0.5\textwidth]{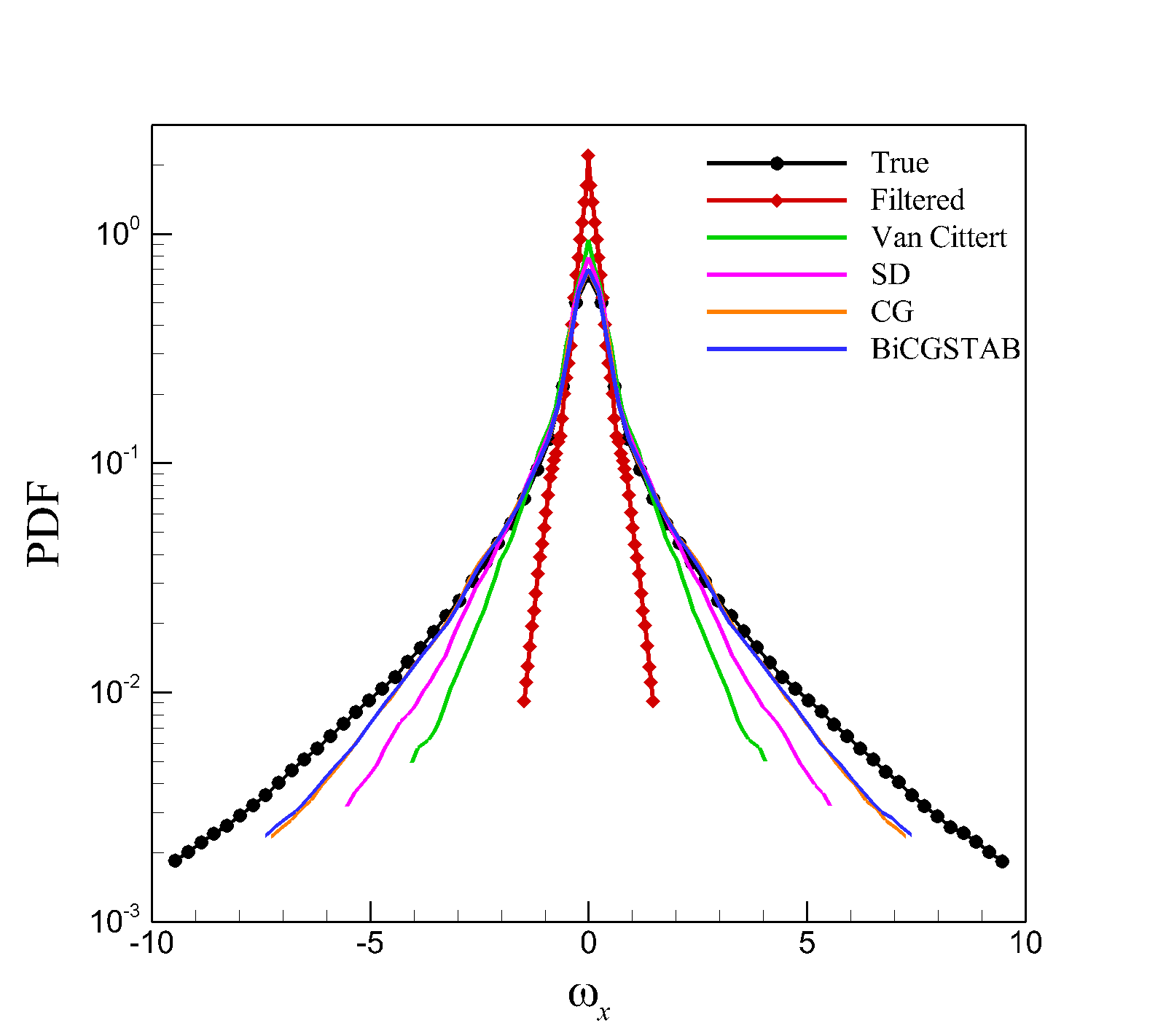}}
\subfigure[BiCGSTAB ($512^3$)]{\includegraphics[width=0.5\textwidth]{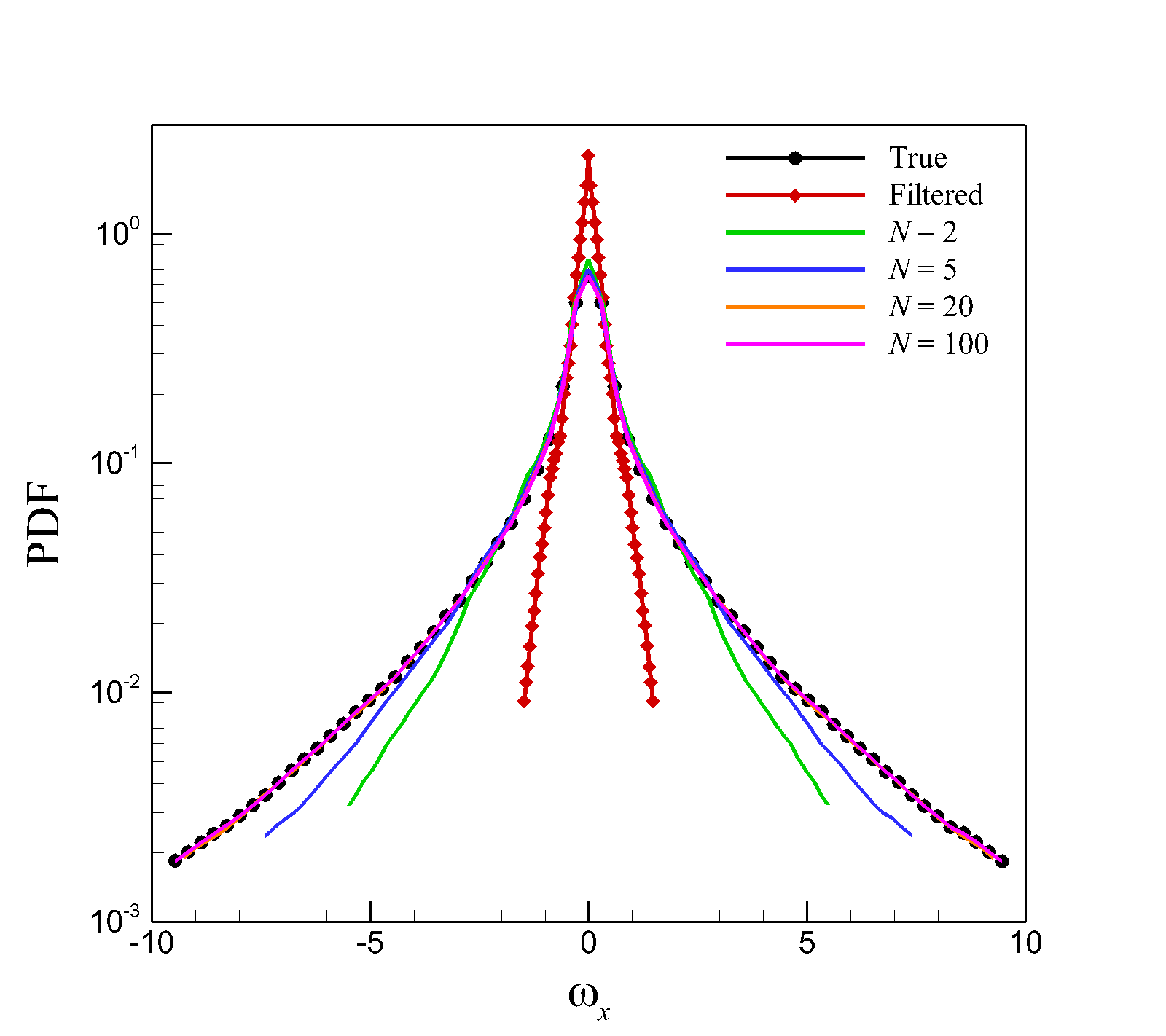}}
}\\
\mbox{
\subfigure[$N=5$ ($64^3$)]{\includegraphics[width=0.5\textwidth]{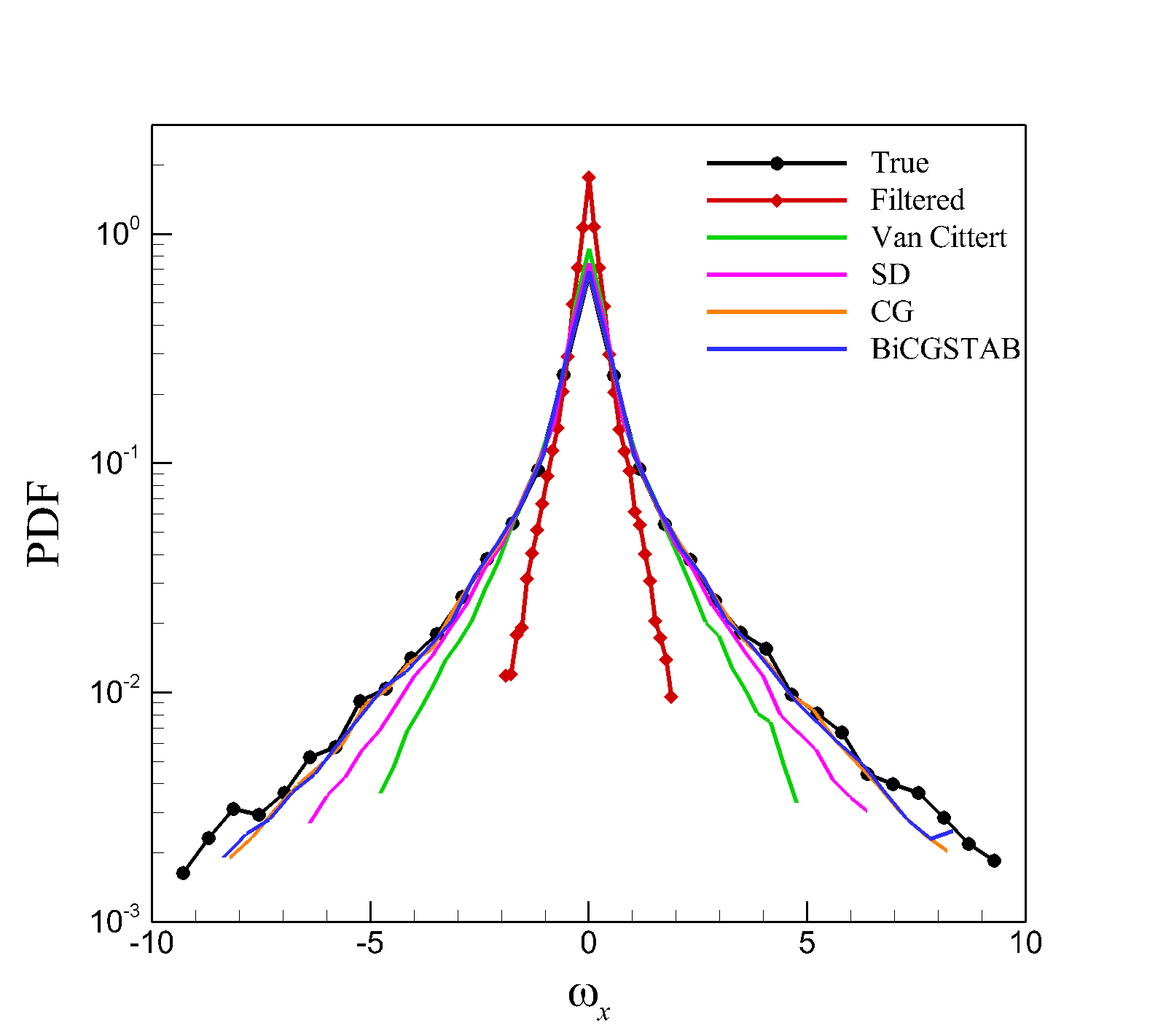}}
\subfigure[BiCGSTAB ($64^3$)]{\includegraphics[width=0.5\textwidth]{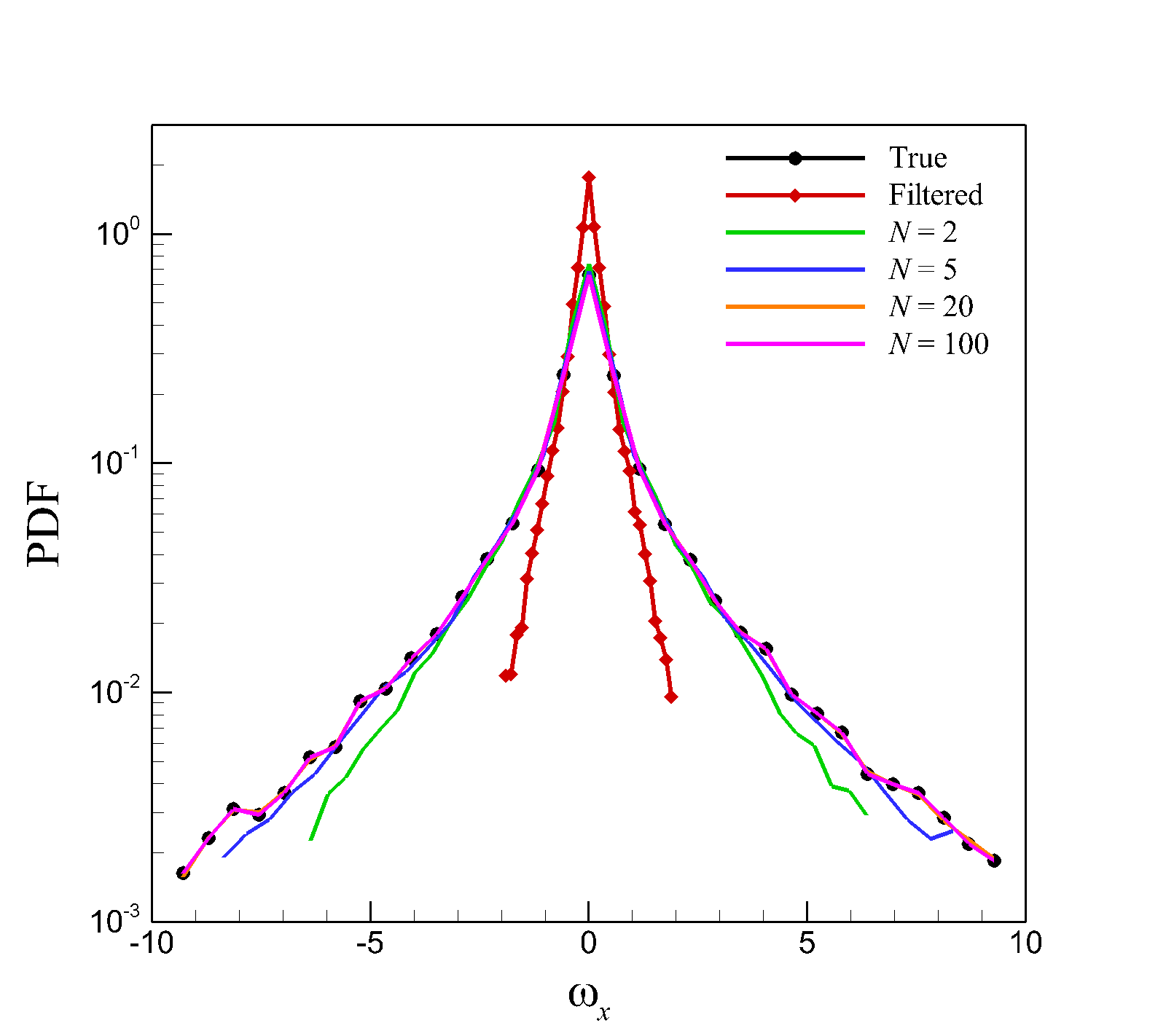}}
}
\caption{Probability density function (PDF) of the three-dimensional turbulence data to illustrate recovery performance of the proposed iterative approaches starting from the filtered input data to predict the true data.}
\label{f:3dpdf}
\end{figure}

To further validate our claims about the potential of the proposed deconvolution procedures, a similar analysis is performed in this section for the three-dimensional Taylor-Green vortex problem \citep{taylor1937mechanism}, which models the decay of homogeneous isotropic turbulent incompressible flow that develops from a single-mode initial condition. The fundamental mechanism is the enhancement of vorticity by vortex stretching and the consequent production of small scale eddies. Energy is transferred from low wave numbers (large scales) to high wave numbers (smaller scales). This energy cascade process controls the turbulent energy dynamics and hence the global structure of the evolution of the turbulent flow. The computational domain used in the present study is a periodic cubic box whose edge has a length of $2\pi$. We note that this flow configuration is perhaps the simplest system in which to study the generation of smaller scale motions and the resulting turbulence. Therefore, this test problem has been extensively used for testing turbulence models. Further details on the problem can be found elsewhere \citep{san2015posteriori}. In a manner similar to the 2D turbulence test case presented earlier, we first generate high fidelity data for $Re=1600$ on a uniform grid with $512^3$ degrees of freedom. After the onset of turbulence, we store a snaphsot of all flow quantities at $t=10$. A set of coarse-grained data for a resolution of $64^3$ grid points is also extracted by using a coarse-injection approach. Our a-priori assessments are based on these two data sets.

Following the same approach provided in our two-dimensional analysis, we use the elliptic differential filter given by Eq.~(\ref{eq:df}) to obtain a filtered flow data and use this filtered data set as initial condition for the proposed approximate deconvolution procedures to test their recovery performance. Similar to the 2D test cases, we use the filtering strength of $\gamma=0.01$ and $\gamma=0.1$ for high-fidelity and coarse-fidelity data sets, respectively. Fig.~\ref{f:3disoN} shows isosurfaces of $x$-component of the vorticity $\omega_x = \pm 0.3$ to illustrate a-priori recovery process of the true data from the filtered data using $N=5$ iterations on a resolution of $512^3$. It can be seen that the conjugate gradient based methods (i.e., both CG and BiCGSTAB) result in a remarkable reconstruction of data. A sensitivity analysis with respect to $N$ is also demonstrated in Fig.~\ref{f:3disoBi} by using the BiCGSTAB method. Residual history of the iterative approximate deconvolution processes are shown in Fig.~\ref{f:3dres} for both high and coarse resolutions. We can make an observation here that all the proposed deconvolution approaches display similar trends witnessed in the 2D test case. The BiCGSTAB method using only one iterate provides more accurate estimates than the recovery obtained by the CG method using two iterates. The BiCGSTAB results are also more accurate than those of obtained by the Van Cittert method using five iterates. This clearly demonstrates the success of the conjugate gradient methods for the AD process.

For symmetric positive definite systems, the BiCG delivers the same results as the CG, but at twice the cost per iteration. Since the BiCGSTAB has smoother convergence, it is normally preferred over the BiCG in the non-Hermitian cases. This convergence property makes it less prone to an inaccurate solution when solving large scale problems, where there exists either a stagnation or strong instability \cite{broyden2004krylov}. In the BiCGSTAB framework (e.g., see Algorithm~\ref{alg:bi}), a constant $\omega_j$ is introduced in each iteration and is chosen as to minimize residual with respect to $\omega_j$. Therefore, the BiCGSTAB can be considered a combination of the BiCG and GMRES approaches. Our results also confirm the positive effect of the conjugate gradient stabilization for the deconvolution process, where its well-posedness is an open problem in AD framework.

Fig.~\ref{f:3dspec} demonstrates our statistical assessments for both the high-fidelity and low-fidelity data sets in term of energy spectra. The ideal Kolmogorov scaling given by $k^{-5/3}$ is also included in each subfigure \citep{frisch1995turbulence}. Firstly, the CG and BiCGSTAB methods lead to a significant enhancement in inertial range recovery for the kinetic energy spectra. As shown in Fig.~\ref{f:3dpdf}, the PDF comparison between true and recovered fields also indicates the ability of the proposed AD procedures for reconstructing the true underlying trends from the filtered input data. Although we only show the probability density function of the $x$-component of the vorticity field, it is noted that the similar trends are also observed from the other flow quantities. The computational cost analysis of the 3D recovery is also presented in Table~\ref{tab:1}, which clearly verifies the superior performance of the proposed conjugate gradient based iterative techniques for AD process. Overall, we may conclude that the CG or BiCGSTAB approximate deconvolution algorithms can be effectively used in structural modeling of turbulence. They offer substantial performance improvement over the standard Van Cittert algorithm which has been used extensively in sub-filter scale modeling of turbulent flows.

\section{Conclusion}
\label{s:con}

In the present work we formulate several Krylov subspace iterative methods for the AD process of sub-filter scale recovery problem of turbulent flows. Compared with the existing traditional Van Cittert iterations for the AD based structural turbulence models, the proposed AD procedures equipped with the conjugate gradient based iterative methods possess potential advantage in terms of efficiency and accuracy. Although both the CG and BiCGSTAB methods result in similar computational performances, which are superior than the standard Van Citert and steepest descent methods, our assessments on recovery of energy spectra and probability density functions demonstrate more stable behavior when the BiCGSTAB method is being used. Our a-priori assessments also suggest that the proposed generalized deconvolution methodology can be used to accelerate LES computations in a-posteriori setting along with a sensitivity study to determining dependence of different filter types, a topic we intend to investigate further in a future study.

\begin{acknowledgements}
The computing for this project was performed by using resources from the High Performance Computing Center (HPCC) at Oklahoma State University.
\end{acknowledgements}

\bibliographystyle{spmpsci}      
\bibliography{References}   

%
%

\end{document}